\documentclass[aps,prl,twocolumn,footinbib,superscriptaddress]{revtex4-2}

\pdfoutput=1

\usepackage[utf8]{inputenc} 
\usepackage{amsmath}
\usepackage{amssymb}
\usepackage{xcolor}
\usepackage{graphicx}
\usepackage{subfigure}
\usepackage{dcolumn}
\usepackage{bm}
\usepackage{microtype}
\usepackage{threeparttable}
\usepackage{mathtools}
\usepackage[colorlinks=true]{hyperref}
\usepackage{physics}
\usepackage{braket}
\usepackage{comment}
\usepackage{soul} 
\usepackage{bbold} 
\usepackage{ragged2e}
\usepackage{graphicx,overpic}
\usepackage[most]{tcolorbox} 
\usepackage{todonotes} 
\usepackage{framed}

\tcbset{highlightstyle/.style={
  colback=yellow!20,
  colframe=yellow!80!black,
  boxrule=0pt,
  arc=2pt,
  outer arc=2pt,
  boxsep=2pt,
  left=2pt,
  right=2pt,
  top=2pt,
  bottom=2pt
}}

\newcommand{\be}{\begin{eqnarray}}
\newcommand{\ee}{\end{eqnarray}}

\renewcommand{\Re}{\operatorname{Re}}

\begin{document}

\title{Entanglement entropy in holographic CFTs with generic boundaries}

\author{Peng-Xiang Hao}
\email{pxhao@himis-sz.cn}
\affiliation{Hetao Institute of Mathematics and Interdisciplinary Sciences, Futian Bonded Area, Shenzhen 518045, China}

\author{Fabio Ori}
\email{Fabio.Ori@UGent.be}
\affiliation{Department of Physics and Astronomy, Ghent University, 9000 Ghent, Belgium}

\begin{abstract}
\noindent Entanglement entropy in holographic conformal field theories with boundaries of generic spacetime signature cannot be described entirely within the standard holographic duality to a real bulk spacetime. We show, both in static and time-dependent setups, that an exact match with the conformal field theory predictions in two boundary dimensions entails an extension of the dual spacetime to complex coordinates. As a consequence, relevant bulk objects such as Ryu-Takayanagi surfaces and branes are in general complex, yet anchored on a real asymptotic boundary preserving the physical meaning of field-theoretical quantities. We also provide a third numerical check of the holographic prediction with a two-dimensional system of free Gaussian fermions exhibiting the same asymptotic regimes across different entanglement phases.
\end{abstract}

\maketitle

\noindent 
\textbf{\emph{Introduction.--}} Boundaries are ubiquitous in quantum systems: examples include physical edges, interfaces, defects. In open quantum systems, boundary effects play an essential role in determining quantum correlations~\cite{Landi:2021dsa}. Most measures of entanglement are likewise significantly affected by the presence of boundaries.

Boundary conformal field theory (BCFT) provides a universal framework for describing quantum field theories in the presence of boundaries \cite{Cardy:1984bb,Cardy:2004hm}. The power of BCFTs in capturing the behavior of quantum systems at critical points is apparent in the context of tensor networks \cite{Orus:2013kga,Cirac:2020obd}, where the scale invariant regime of a network with a boundary has been matched to BCFT predictions \cite{Iino:2019vxd,Iino:2019rxt,Iino:2020ipa}.

A primary tool to quantify boundary effects in theories with a holographic dual is the Ryu--Takayanagi~(RT) formula~\cite{Ryu:2006bv,Hubeny:2007xt}, showing that the entanglement entropy of a boundary region in a holographic field theory can be computed geometrically through the area of extremal surfaces in the dual gravitational theory.
In the AdS/BCFT correspondence, the boundary of the field theory is related to an end-of-the-world (EOW) brane extending in the bulk and anchored on the boundary itself~\cite{Takayanagi:2011zk,Fujita:2011fp,Nozaki:2012qd}.

The conventional formulation of the holographic correspondence between Anti-de Sitter (AdS) spacetime and BCFTs (AdS/BCFT) is mainly developed for timelike boundaries, for which the EOW brane and the extremal RT surfaces admit a description within a real Lorentzian bulk geometry. However, more general causal structures of EOW branes naturally arise in holography. Spacelike branes can appear as state-preparation or final-state boundary conditions, where Euclidean path-integral constructions are followed by Lorentzian evolution \cite{Akal:2021dqt}. They also arise in generalized AdS/BCFT constructions involving supercritical branes and de Sitter (dS) regions \cite{Akal:2020wfl}, as well as holographic descriptions of systems with null boundaries and flat space holography \cite{Hao:2025ocu}.

These developments suggest that the standard real Lorentzian AdS/BCFT prescription should be reconsidered when the causal structure of the boundary is changed. Inconsistencies with the traditional extremal-surface prescription arise, such as extremality leading to self-intersecting RT surfaces~\cite{Hao:2024nhd}.
Similar puzzles raise the question of providing a more general framework to consistently describe CFTs with generic boundaries.

In this work, we argue that a consistent description of CFTs with generic boundaries requires extending the bulk spacetime to complex coordinates. The emergence of a complex bulk geometry is not an artifact of a particular analytic continuation, but rather a general feature of the holographic duality, as it crucially affects all relevant bulk objects with a boundary interpretation: not only RT surfaces, but also EOW branes. The holographic entanglement (pseudo) entropy is then obtained from complex extremal surfaces whose endpoints lie on a complexified brane. The dominant saddle is selected by minimizing the real part of their area, together with time-ordering prescriptions~\cite{Bernamonti:2026pxo}. Most importantly, the complexified bulk remains anchored on a real asymptotic boundary, preserving the physical interpretation of the BCFT.

We illustrate our proposal via different instances of the AdS$_3$/BCFT$_2$ correspondence: static spacelike boundaries, finite-cutoff branes supporting lower-dimensional gravitational dynamics and time-dependent boundaries. The extremization of the geodesic length functional directly leads to complex saddle points on the analytically continued EOW brane. The resulting complex geodesic lengths reproduce the entanglement (pseudo) entropy obtained from correlation functions of twist operators in the BCFT. In the static case, the asymptotic behavior of the entropy for small or large regions can also be reproduced by a two-dimensional theory of Gaussian free fermions.

Complex bulk saddles have also appeared in pseudo entanglement entropy \cite{Doi:2022iyj,Doi:2023zaf}, a complex extension of entanglement entropy associated to non-unitary transition matrices. The natural geometric interpretation of its holographic dual is also in terms of complex extremal surfaces \cite{Heller:2024whi,Heller:2025kvp}. Our AdS/BCFT results can then be seen as a transition in the boundary region size from (real) entanglement entropy to (complex) pseudo entropy. More in general, they suggest that complex bulk configurations represent a broader feature of holography whenever real ones fail to capture the relevant boundary observables.

\noindent
\textbf{\emph{Proposal.--}} Consider a codimension-1 boundary region $A$ in a holographic BCFT$_d$ dual to a spacetime manifold $\mathcal{M}_{d+1}$, and denote by $B_{d-1}$ the boundary of the BCFT, whose signature can in general be time-dependent.

\begin{figure}
    \centering
    \includegraphics[width=\linewidth]{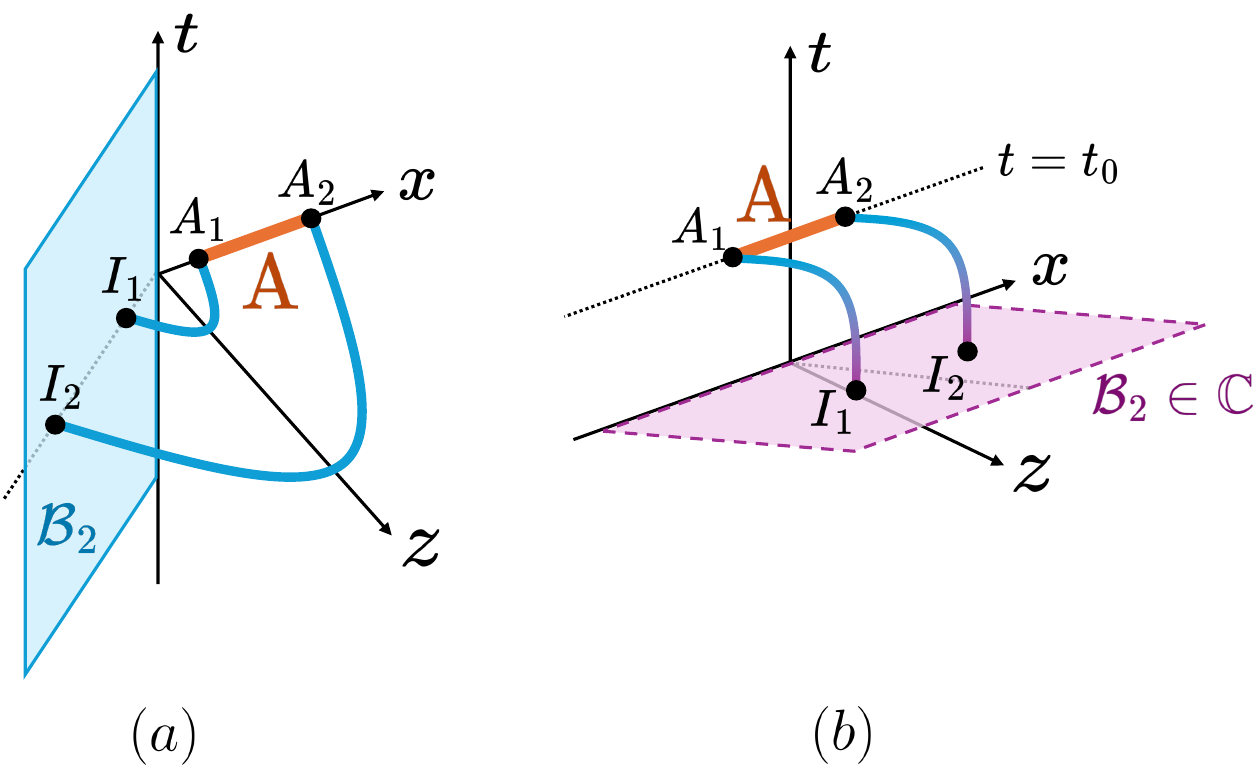}
    \caption{Pictorial representation of the disconnected contribution to the entanglement (pseudo) entropy of a spacelike region $A=[A_1,A_2]$ in the AdS$_3$/BCFT$_2$ correspondence. (\textit{a}) For a timelike brane, the RT surface (blue) is real and anchored at two points $I_1$, $I_2$ on the real brane $\mathcal{B}_2$. (\textit{b}) When the brane is spacelike, the anchoring points $I_1$, $I_2$ that satisfy the RT condition are complex. As a consequence, the RT surface is complex (purple), yet still anchored at a real boundary.}
    \label{fig:proposal}
\end{figure}

Let us briefly recall the holographic entanglement entropy prescription in the AdS/BCFT correspondence, where two competing channels exist~\cite{Takayanagi:2011zk,Fujita:2011fp}. 
The \emph{connected} phase is the area of the minimal-area, bulk extremal surface $\gamma_{A}$ anchored at the boundary of $A$:
\begin{equation}
S_A^{\rm con} = \min_{\gamma_A}\frac{\mathrm{Area}(\gamma_{A})}{4G_N}\ .
\end{equation}
The \emph{disconnected} phase involves instead a (possibly non-connected) extremal surface $\gamma_{A,\,I}$ ending  at a region $I$ on the EOW brane $\mathcal{B}_d$, known as the \emph{island}: see Fig.~\ref{fig:proposal}\,(\textit{a}) for a three-dimensional example. The island location is fixed by extremizing the total area of $\gamma_{A,\,I}$:
\begin{equation}
S_A^{\rm dis} =
\underset{I\, \subset\,  \mathcal{B}_d}{\mathrm{ext}} 
\frac{\mathrm{Area}(\gamma_{A,\, I})}{4G_N}\ .
\end{equation}
This extremum is a saddle point for the area functional. The resulting holographic entanglement entropy is then the minimum of (the real part of) the two contributions,
\begin{equation}
S_A = \min_{\operatorname{Re} S_A} \big\{ S_A^{\rm con},\; S_A^{\rm dis} \big\}\,,
\end{equation}
where all RT surfaces areas are intended to be consistently regularized by a UV cutoff near the asymptotic boundary.
\\

We propose that the disconnected RT surfaces dual to the entanglement (pseudo) entropy with boundaries of any spacelike signature are in general anchored at a \emph{complex} bulk brane $\mathcal{B}_d$ dual to $B_{d-1}$. More precisely, $\mathcal{B}_d$ lives in a complexified extension of the spacetime $\mathcal{M}_{d+1}$ obtained by analytically continuing the coordinates,
\begin{align}
    &\mathcal{M}_{d+1}(x^\mu\in \mathbb{R})\rightarrow \mathcal{M}_{d+1}(x^\mu\in \mathbb{C})\\[.5em]
    &\mathcal{B}_d(x^\mu\in\mathbb{R})\rightarrow \mathcal{B}_d(x^\mu\in \mathbb{C})\,,
\end{align}
such that $\mathcal{B}$ is still anchored to the \emph{real} spacetime boundary $B$, i.e.\ $\mathcal{B}|_{\partial\mathcal{M}}=B$. This boundary condition is crucial as it preserves the physical interpretation of the BCFT.

The (pseudo) entropy for the disconnected configuration is then given by generically complex RT surfaces with one end anchored at the boundary of $A$, and the other at the island $I \subset \mathcal{B}_d$. The location of $I$ is still fixed by extremization of the surface area functional. If all saddle points have areas with different real parts, the entanglement entropy will be defined by the minimal one,
\begin{equation}\label{eq:RT_BCFT}
    S_A=\min_{\Re \operatorname{Area}(\gamma_{A,\,I})}\frac{\operatorname{Area}(\gamma_{A,\,I})}{4G_N}\ .
\end{equation}
In some cases, if multiple saddle points with the same real part exist, time ordering of the endpoints of $A$ can select the dominant RT saddle  \cite{Bernamonti:2026pxo}, as shown later.

\noindent
\textbf{\emph{Static branes.--}}
We illustrate the proposal with two static-brane configurations in AdS$_3$/BCFT$_2$, showing how the complex nature of the EOW brane emerges when evaluating the RT surfaces. The resulting complex (pseudo) entropy matches the dual BCFT$_2$ prediction.
\\

\noindent\emph{1. Constant-time brane in Poincar\'e AdS$_3$.}\;\;
The metric is $ds^2 = z^{-2}(-dt^2+dx^2+dz^2)$, where $z=r^{-1}$ is the holographic direction, and $\mathcal{B}_2$ is the plane $t=0$. The spacelike boundary can be interpreted as state preparation: the vacuum state of the CFT$_2$ is created by adding layers of entanglement to an IR disentangled state as in tensor networks \cite{Akal:2020wfl}, and is then evolved in real time as $t>0$~\footnote{The authors of~\cite{Akal:2020wfl} studied the holographic dual of a BCFT with a spacelike boundary. However, their approach differs from the one considered in this Letter, as in their case a real Lorentzian brane $\tau=\lambda z$ describing a spacelike CFT boundary emerges only after a double analytic continuation: first, a Wick rotation $\tau=it$ making the brane effectively complex, then a continuation of $\lambda=i\tilde \lambda$, $\tilde\lambda\in\mathbb{R}$, turning it to real again at the expense of changing the reality of the boundary conditions.}.
We consider a spacelike interval of size $L$ on the $t=t_0$ time-slice with endpoints $A_1=(t_1,x_1,z_1)=(t_0,x_1,\epsilon)$ and $A_2=(t_2,x_2,z_2) =(t_0,x_1+L,\epsilon)$, where $t_0>0$, $L>0$ and $\epsilon$ is a regulator. See Fig.~\ref{fig:proposal}\,(\textit{b}) for a visualization.

\begin{figure}[tb]
  \centering
  \includegraphics[width=.95\columnwidth/2]{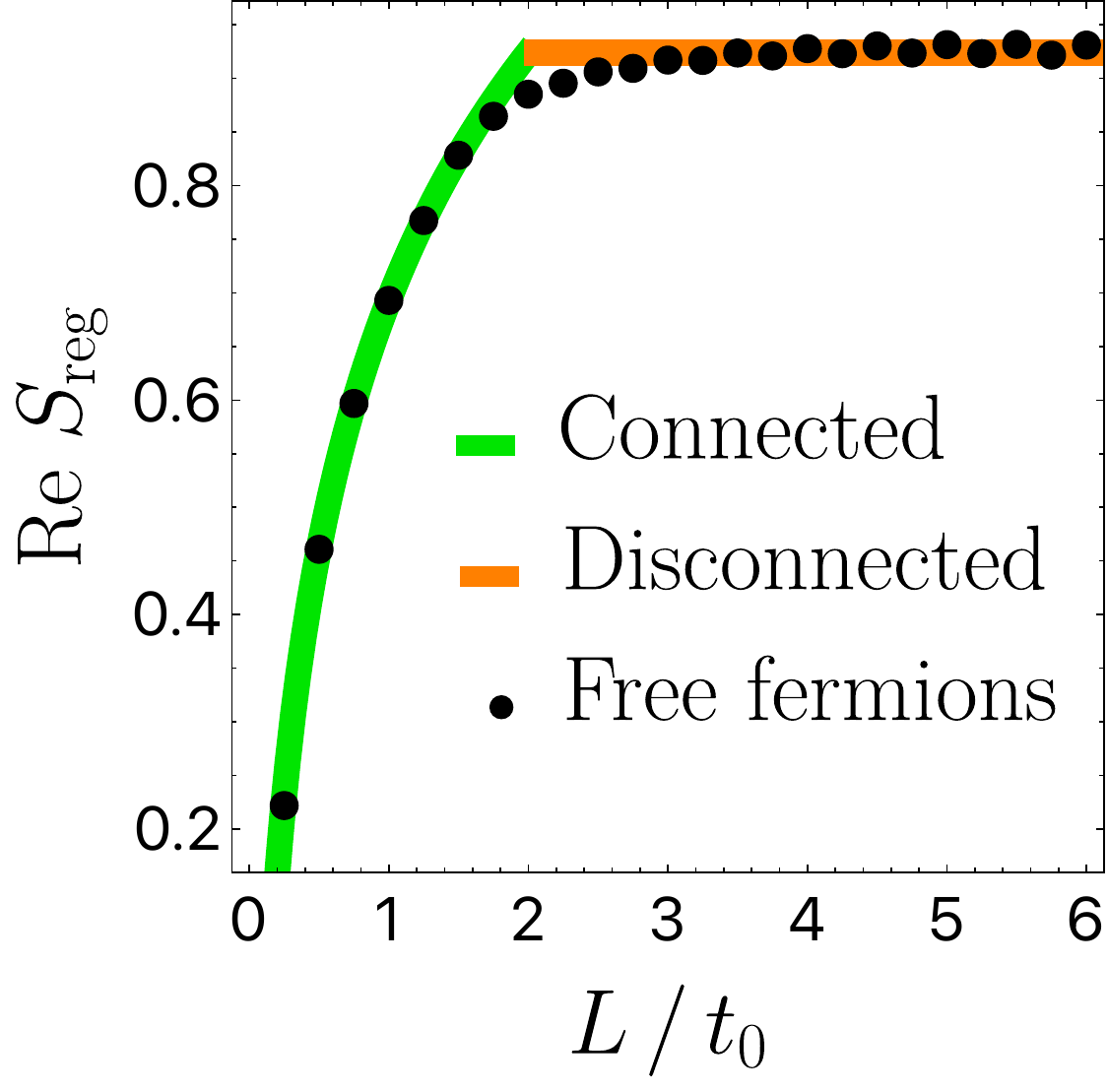}
  \hspace{.02cm}
  \includegraphics[width=.95\columnwidth/2]{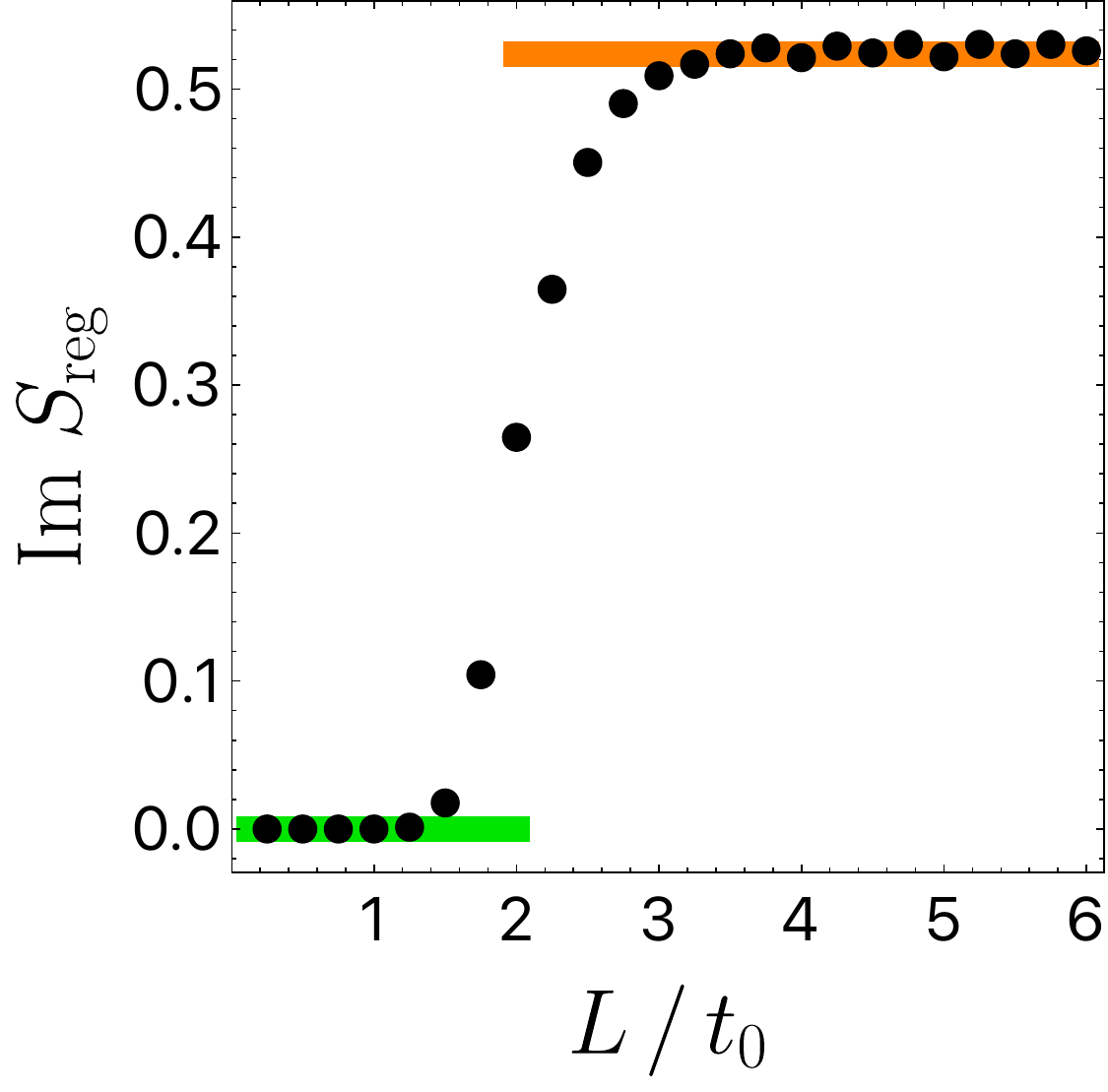}
  \caption{The entanglement (pseudo) entropy of a spacelike interval of size $L$ lying on the time-slice $t=t_0$ for the constant-time brane $t=0$ in Poincaré AdS$_3$. We plot the regularized quantity $S_\mathrm{reg}\equiv \frac{6}{c}S-\log\epsilon$ as a function of the dimensionless ratio $L/t_0$. Both the real and imaginary parts of the entropy display a phase transition from the connected \eqref{eq:Dcon_case1} to the disconnected~\eqref{eq:Ddisc_case1} configuration as $L/t_0$ increases. The holographic result (solid lines) is overlaid with an evaluation of the same quantity for (1+1)-dimensional Gaussian free fermions, which exhibits the same asymptotic regimes as $L\ll 2t_0$ and $L\gg 2t_0$.}
  \label{fig:con_t_ee}
\end{figure}

The connected configuration has geodesic length
\begin{equation}\label{eq:Dcon_case1}
    D_{\rm con} = 2\log\frac{L}{\epsilon}\ .
\end{equation}
In the disconnected phase, instead, two geodesics connect the endpoints $A_1$, $A_2$ to the points $I_1=(0,x_{I1},z_{I1})$ and $I_2=(0,x_{I2},z_{I2})$ on the brane, see again Fig.~\ref{fig:proposal}~(\textit{b}). Extremizing the length functional
\begin{equation}
    \mathrm{Len}=\int_\gamma d\lambda\ \sqrt{\frac{-\dot t^2(\lambda)+\dot x^2(\lambda)+\dot z^2(\lambda)}{z^2(\lambda)}}
\end{equation}
along a trajectory $\gamma:(t(\lambda),x(\lambda),z(\lambda))$ with a parameter $\lambda$ that satisfies the appropriate boundary conditions, gives $x_{I1} = x_1$, $x_{I2}=x_2$ and $z_{I1} = z_{I2} = i t_0$. There is a sign multiplicity in the imaginary part of $z_{I1}$, $z_{I2}$ and consequently in the length: it can be solved by requiring time ordering of the endpoints with respect to the brane, along the lines of \cite{Bernamonti:2026pxo}:  more details can be found in the Supplemental Material. As the saddle point appears at a \emph{complex} $z$-coordinate, the anchor point necessarily lies on the analytic continuation of the brane $\mathcal{B}_2(t=0,\,z\in\mathbb{C})$ to complex values. Substituting back, the length is
\begin{equation}\label{eq:Ddisc_case1}
D_{\rm disc} = 2\log\left(\frac{2t_0}{\epsilon}\right) +i\pi \,.
\end{equation}
The dominant (pseudo) entropy contribution as a function of the interval size $L$ is given by \eqref{eq:RT_BCFT} and is plotted in Fig.~\ref{fig:con_t_ee}. The real parts of the connected (green) and disconnected (orange) configurations cross at $L = 2t_0$. Correspondingly, the imaginary part undergoes a first‑order transition.

\vspace{4pt}
\noindent\emph{Boundary dual.}\;\;
In the dual BCFT$_2$, the entanglement entropy is extracted from twist-anti-twist correlators \cite{Calabrese:2009qy},
\begin{equation}\label{eq:ent_twist_correlator}
    S_A=\lim_{n\to 1}\frac{1}{1-n}\log{\braket{\Psi|\sigma_n(t_1,x_1)\,\tilde\sigma_n(t_2,x_2)|\Psi}}\,.
\end{equation}
We follow the approach of \cite{Bernamonti:2026pxo} to unambiguously define the entropy across time-slices by analytically continuing the twist and anti-twist insertions in the complex plane while preserving time-ordering. As the connected configuration is just the usual CFT$_2$ prediction reproducing the bulk length \eqref{eq:Dcon_case1}, we will focus on the disconnected one.

In Euclidean signature $\tau =it$, with the boundary at $\tau=0$ and the BCFT$_2$ defined in the upper half‑plane $\tau>0$, twist and anti-twist operators $\sigma_n$, $\tilde\sigma_n$ of scaling dimension $\Delta_n=\frac{c}{12}\left(n-\frac{1}{n}\right)$ have the one‑point function 
\begin{equation}\label{eq:onept_eucl}
\langle \sigma_n(\tau,x) \rangle_{\mathrm{E}} =\langle \tilde\sigma_n(\tau,x) \rangle_{\mathrm{E}}= A_n(2\tau)^{-\Delta_n}
\end{equation}
with a real constant $A_n$ fixed by the boundary conditions. The factorized boundary channel then gives \cite{McAvity:1995zd,DiFrancesco:1997nk}
\begin{equation}
    \langle \sigma_n(\tau_1,x_1)\,\tilde\sigma_n(\tau_2,x_2) \rangle_{\mathrm{E,\,disc}} \sim \frac{A_n^2}{(4\tau_1\tau_2)^{\Delta_n}}\,.
\end{equation}
A Wick rotation that preserves time-ordering can be implemented by an $i\varepsilon$-prescription as $\tau\to it+\varepsilon$, see \cite{Bernamonti:2026pxo}. In the small regulator limit $\varepsilon\to 0$, the boundary‑channel two‑point function becomes then
\begin{equation}
\langle T\left\{\sigma_n(t_1,x_1)\,\tilde\sigma_n(t_2,x_2)\right\} \rangle_{\rm disc} \sim \frac{A_n^2}{(e^{i\pi}\,4t_1t_2)^{\Delta_n}}\,.
\end{equation}
Using \eqref{eq:ent_twist_correlator}, this yields the entanglement pseudo entropy in the disconnected channel,
\begin{equation}\label{eq:Sdisc_bdy}
    S_A^{\rm disc}=\frac{c}{3}\log\frac{2t_0}{\epsilon}+i\frac{\pi c}{6}+S_{\rm bdy},
\end{equation}
where $S_{\rm bdy}$ is a contribution from the boundary entropy that vanishes in the case of the $t=0$ brane. In the Supplemental material we show how it can be reproduced from the bulk by considering more general brane shapes. The remaining part is exactly matched by the disconnected geodesic length~\eqref{eq:Ddisc_case1} with $c=\frac{3}{2G_N}$. This agreement supports the necessity of analytically continuing the brane to complex coordinates to capture the BCFT result.

As a cross-check, the entanglement (pseudo) entropy of the region $A$ can also be computed in a (1+1)-dimensional system of Gaussian free fermions~\cite{Casini:2009sr} with a simulated spacelike boundary at $t=0$, using techniques similar to the ones employed for entanglement in time \cite{Milekhin:2025ycm}. Both the real and imaginary parts match the holographic and BCFT predictions in the two asymptotic regimes of small ($L\ll 2t_0$) and large ($L\gg 2t_0$) entangling regions $A$, see Fig.~\ref{fig:con_t_ee} and the Supplemental Material for details.
\\

\begin{figure}[b]
    \centering
    \includegraphics[width=\linewidth]{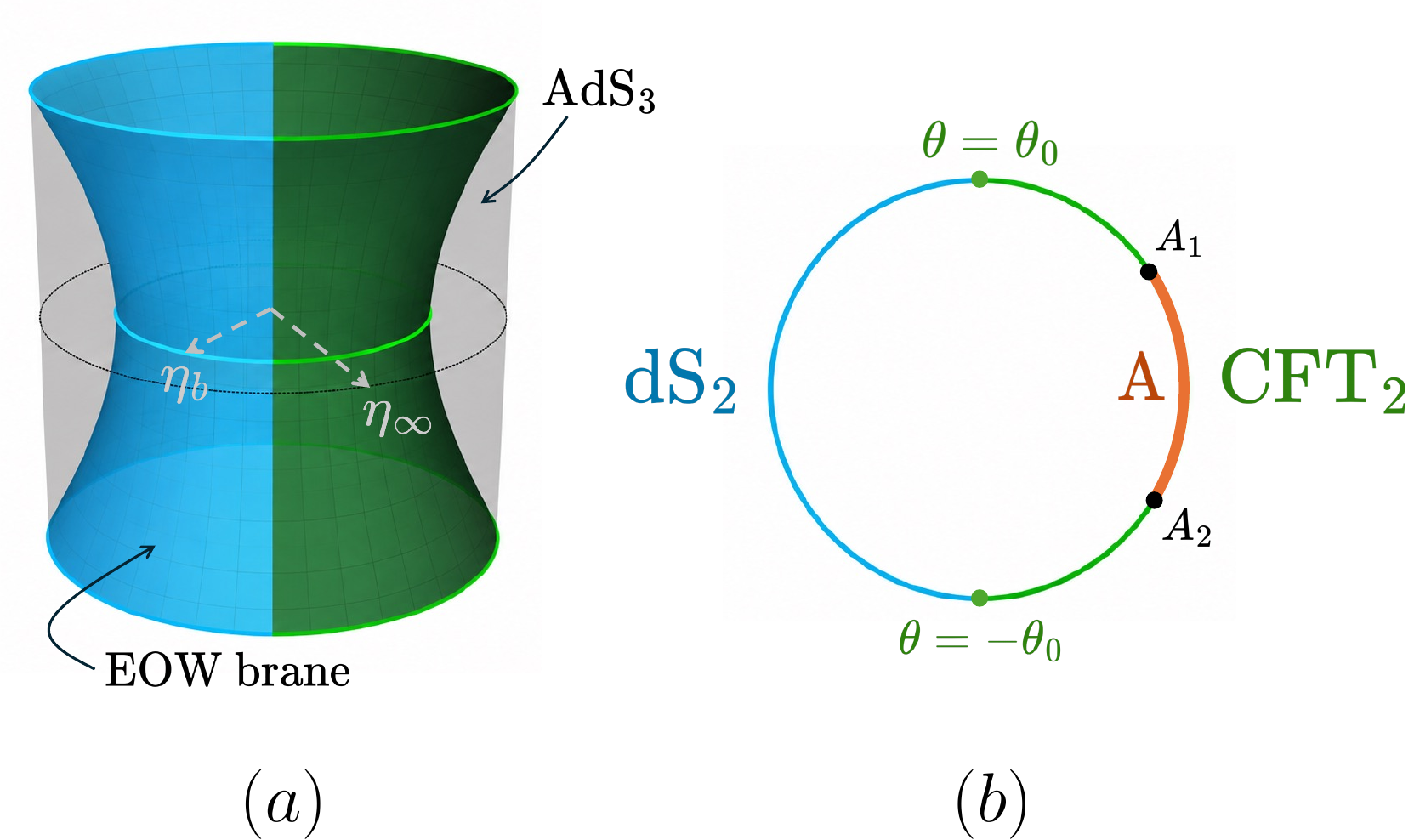}
    \caption{Schematic picture of the AdS$_3$/BCFT$_2$ correspondence in presence of a de Sitter brane at finite cutoff. (\textit{a}) The AdS$_3$ bulk spacetime is sliced in dS$_2$ slices along the $\eta$-coordinate, with the asymptotic boundary at $\eta=\eta_\infty$ and the EOW brane at $\eta=\eta_b$. (\textit{b}) Representation of the constant time-slice where the entangling region $A$ lies. As a consequence of the slicing, the CFT$_2$ is coupled to a dS$_2$ spacetime with dynamical gravity at some value of the angular coordinate $\theta=\pm \theta_0$. The figure is inspired by a similar one appearing in \cite{Hao:2024nhd}.}
    \label{fig:dS_brane}
\end{figure}

\noindent\emph{2.~de Sitter brane with finite cutoff.}\;\;
The bulk is again empty AdS$_3$, but written in an AdS$_3$/dS$_2$ slicing,
\begin{align}
  ds^2 = d\eta^2 + \sinh^2\!\eta\,(-dt^2 + \cosh^2\!t\,d\theta^2)\,,
\end{align}
with the asymptotic boundary at $\eta=\eta_\infty\sim \log\frac{2}{\epsilon}$ and the brane $\mathcal{B}_2$ located at a finite radius $\eta=\eta_b>0$ carrying dynamical dS$_2$ gravity \cite{Hao:2024nhd}, see Fig.~\ref{fig:dS_brane}\,(\textit{a}).

\begin{figure}
    \centering
    \includegraphics[width=.99\columnwidth/2]{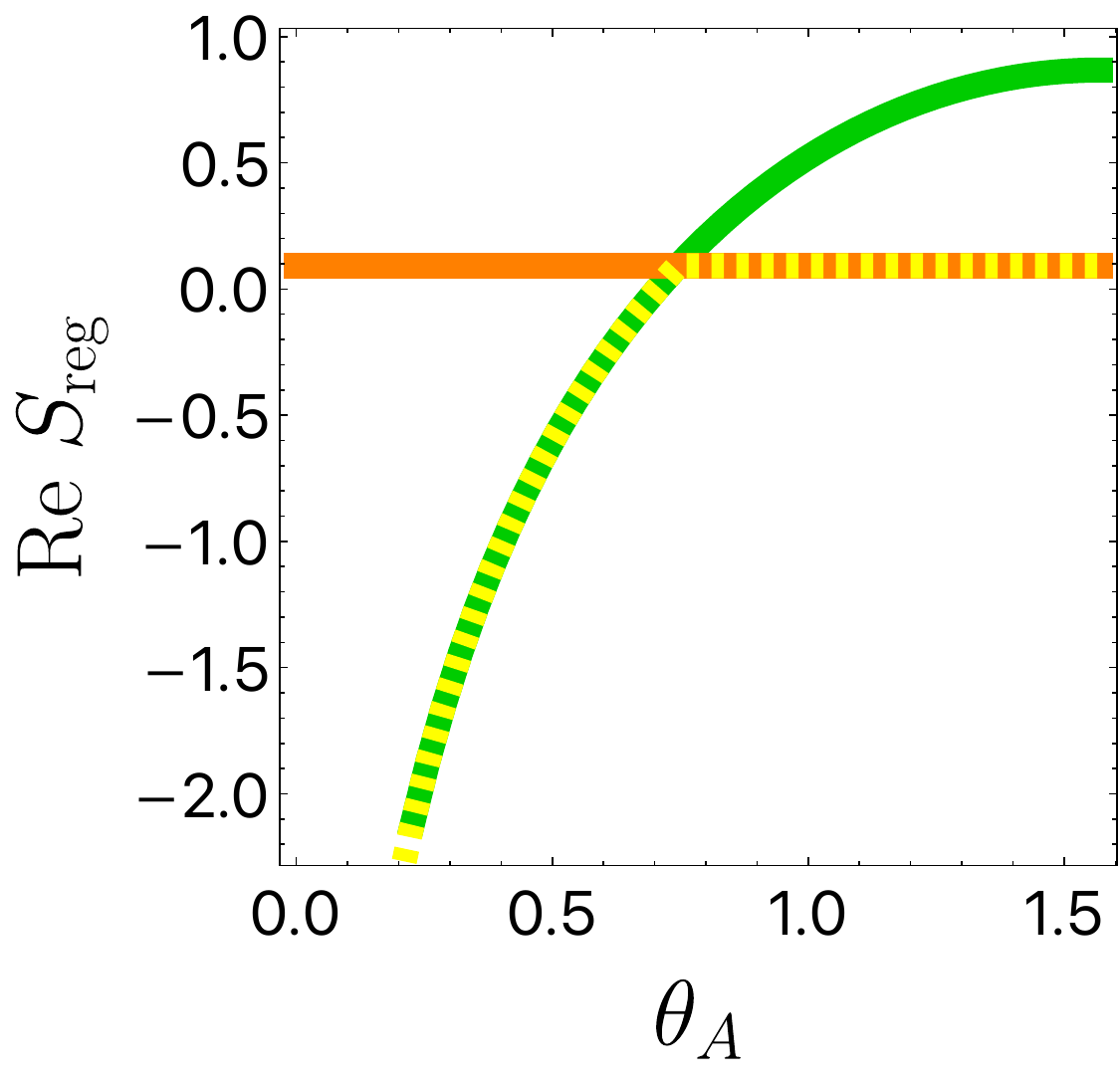}
    \hspace{.02cm}
    \includegraphics[width=.95\columnwidth/2]{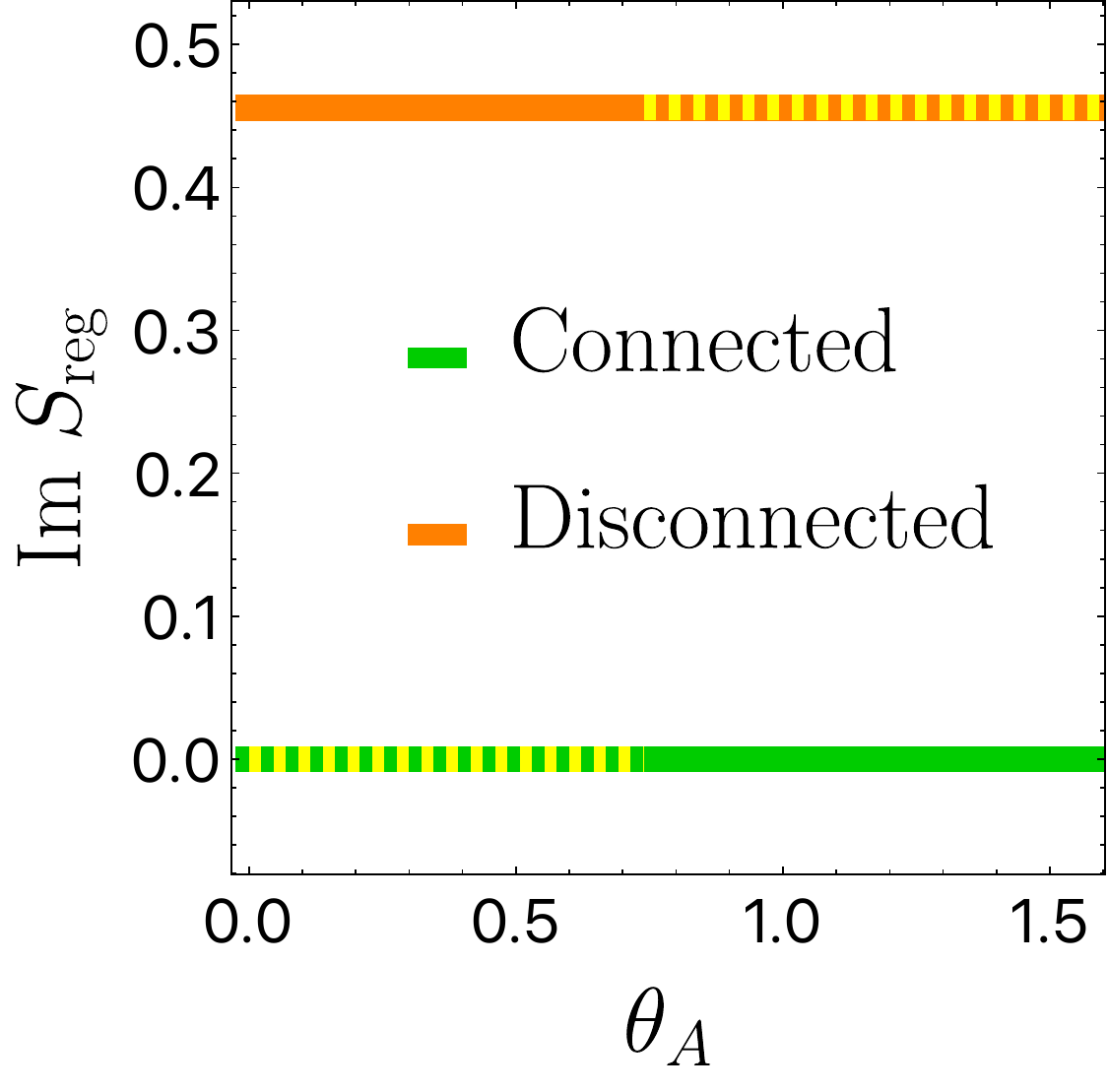}
    
    \caption{Real and imaginary part of the entanglement (pseudo) entropy with a de Sitter brane at finite cutoff as a function of the angular size $2\theta_A$ of the spacelike boundary region $A$. The entropy is regularized as $S_{\mathrm{reg}}\equiv \frac{6}{c}S-2 \eta_\infty$. Here, $t_A=1$, $\eta_b=\frac{1}{4}$, $\theta_0=\frac{\pi}{2}$, and we show the contribution from the disconnected sector \eqref{eq:S_disc_ds} arising for even values of $m$ (odd values have the same real part and opposite imaginary ones in the disconnected phase). The yellow dashed line indicates the dominant contribution according to \eqref{eq:RT_BCFT}.}
    \label{fig:S_dS}
\end{figure}

For endpoints $A_1=(t_1,\theta_1,\eta_1)=(t_A,-\theta_A,\eta_\infty)$ and $A_2=(t_2,\theta_2,\eta_2)=(t_A,\theta_A,\eta_\infty)$, the connected geodesic length is
\begin{align}
  D_{\rm con} = 2\eta_\infty + 2\log \bigl(\cosh t_A\sin\theta_A\bigr)\,.
\end{align}
In the disconnected or island phase, a geodesic connects $A_1$ to a point $I_1=(t_I,\theta_I,\eta_b)$ on $\mathcal{B}_2$, and similarly for $A_2$. 
Extremization of the length functional over $t_I,\theta_I$ yields the family of complex saddles
\begin{equation}\label{eq:complex_saddles_dS}
  \theta_{I1,\,2} = \mp \theta_A + \frac{\pi}{2} + \pi k,\qquad
  t_{I1,\,2} = i\pi\Bigl(m+\frac12\Bigr)\,,
\end{equation}
where $k,m\in\mathbb{Z}$. They clearly lie on the complexified brane $\mathcal{B}_2(\eta=\eta_b,\,t\in\mathbb{C},\,\theta\in\mathbb{R})$ and have length
\begin{equation}
\begin{aligned}
  D_{\mathrm{disc}} = 2 \eta_\infty +2\log&\left[\cosh\eta_b\right.\\
  & \left. +\, i(-1)^m \sinh\eta_b \sinh t_A\right],
\end{aligned}
\end{equation}
which is generically complex for any finite $\eta_b$. From the proposal \eqref{eq:RT_BCFT}, the entanglement pseudo entropy is then
\begin{equation}
\begin{aligned} \label{eq:S_disc_ds}
    S_A^{\mathrm{disc}} = \min_{m\in \mathbb{Z}}\, \frac{c}{3}&\left\{\eta_\infty+\log\left[\cosh\eta_b\right.\right.\\
    &\left.\left. + i(-1)^m \sinh\eta_b \sinh t_A\right]\right\},
\end{aligned}
\end{equation}
where the minimization is understood to act on the real part only. Fig.~\ref{fig:S_dS} illustrates the exchange of dominance between the two contributions for the class $m$ even.

As detailed in the Supplemental Material, there are also additional real saddles: however, they are unphysical, since they correspond to local maxima of the entropy functional and do not contribute to the Euclidean path integral. Moreover, their dual is ill-defined, as the corresponding RT surface self-intersects in the middle of the spacetime~\cite{Hao:2024nhd}. Only the family of saddles \eqref{eq:complex_saddles_dS} is then allowed, necessarily leading to complex branes.

\vspace{4pt}
\noindent\emph{Double holography.}\;\;  
We can now understand the above result from the perspective of double holography, where the lower‑dimensional quantum gravitational system is realized as an EOW brane embedded in a higher‑dimensional AdS spacetime. In the present setup, we embed the dS$_2$ brane into an AdS$_3$ bulk using the AdS$_3$/dS$_2$ slicing, with the brane located at a constant radial coordinate $\eta=\eta_b$. The gravitational region on the dS$_2$ brane is described by imposing Neumann boundary conditions on a portion of the circle, while the non‑gravitational bath region obeys Dirichlet boundary conditions. In the limit $\eta_b\to \infty$, such inhomogeneous boundary conditions define an AdS$_3$/BCFT$_2$ model equivalent to dS$_2$ gravity coupled to a CFT$_2$ bath. Three equivalent descriptions coexist: the bulk AdS$_3$ with a dS$_2$ EOW brane, the brane perspective of induced gravity on dS$_2$ coupled to the CFT$_2$, and the boundary BCFT$_2$: see Fig.~\ref{fig:dS_brane}\,(\textit{b}) and \cite{Hao:2024nhd}.

We are interested in evaluating the entanglement entropy of a boundary interval $A$ on a constant time-slice in the CFT$_2$ bath. The saddles \eqref{eq:complex_saddles_dS}, leading to the entropy~\eqref{eq:S_disc_ds}, are consistent with the island formula \cite{Penington:2019npb,Almheiri:2019psf}
\begin{equation}\label{eq:island_formula}
S_I(\hat{\rho}_{A}) = \min_I \operatorname*{ext}_{I} \left[ \frac{\mathrm{Area}(\partial I)}{4G_N} + S_{\mathrm{matter}}(A \cup I) \right],
\end{equation}
where $I$ denotes the island with boundary $\partial I$. The first term is the gravitational entropy, which is constant in this case \cite{Hao:2024nhd}, while $S_\mathrm{matter}$ is the matter entropy, i.e.\ the entanglement entropy in the CFT on the global dS$_2$ background.
As shown in the Supplemental Material, extremization over the position of the island $(t_I,\theta_I)$ reproduces the saddles~\eqref{eq:complex_saddles_dS} as $\eta_b\to\infty$. The complex geodesic picture captures the fine‑grained entropy of the bath, and the finite‑cutoff brane must be analytically continued to complex coordinates to support the island. An alternative construction, also leading to complex saddles, has been recently pointed out \cite{Hoback:2026yqj}, as well.

\vspace{4pt}
In summary, both examples illustrate how the extremality of RT surfaces forces the EOW brane to extend into a complexified bulk. The resulting complex saddles reproduce the BCFT or double-holography predictions and are compatible with free field simulations.
\\

\noindent
\textbf{\emph{Time-dependent branes.--}}
The EOW brane complexification we found in static setups extends naturally to time‑dependent backgrounds. A rich class of examples is provided by (1+1)-dimensional moving mirrors \cite{Davies:1976hi,Cardy:2004hm}, i.e.\ CFTs defined on a spacetime with a time-dependent boundary which acts as a `mirror trajectory'.

\vspace{4pt}
\noindent\emph{Holographic moving mirrors.}\;
Any timelike mirror profile in a CFT$_2$ on the half‑plane can be mapped to the static boundary $\tilde{x}=0$ by a chiral conformal transformation. This map lifts to the bulk, yielding an AdS/BCFT dual with a dynamical EOW brane \cite{Akal:2020twv,Akal:2021foz,Akal:2022qei}. In light‑cone coordinates $(u,v)=(t-x,t+x)$ the boundary metric is $ds^2=-du\,dv$. Consider the conformal map
\begin{equation}\label{eq:map_mirrors_bdy}
\tilde u = p(u),\qquad \tilde v = v,
\qquad p'(u)>0,
\end{equation}
which sends the static mirror $\tilde{u}=\tilde{v}$ to the time‑dependent boundary $v=p(u)$. The physical region is $v>p(u)$. The AdS/BCFT construction can be carried out by applying the bulk coordinate transformation dual to \eqref{eq:map_mirrors_bdy},
\begin{equation}
U = p(u),\quad
V = v + \frac{p''(u)}{2p'(u)}\,z^2,\quad
\eta = z\sqrt{p'(u)}\,,
\end{equation}
bridging global AdS$_3$ with coordinates $(U,V,\eta)$
to the static half‑space Poincaré AdS$_3$ with $(u,v,z)$. The resulting EOW brane, originally at $X=\lambda\eta$, is now \cite{Akal:2020twv}
\begin{equation}
v_{\mathrm{brane}} = -\frac{p''(u)\,z^2}{2p'(u)} + p(u) - 2\lambda z\sqrt{p'(u)}\, .
\end{equation}
In the following discussion, we take $\lambda=0$ for simplicity. For a boundary interval $A$ with endpoints $A_1=(u_1,v_1,\epsilon)$ and $A_2=(u_2,v_2,\epsilon)$, the entropy of the disconnected configuration is given by the sum of the length of two geodesics, each connecting the brane to a boundary point. Their regularized length is
\begin{equation}
 D_{\mathrm{disc}}(u_i,v_i) = \sum_{i=1}^2\log\frac{v_i - p(u_i)}{\epsilon\sqrt{p'(u_i)}}\ .
\end{equation}
Its exponential directly yields the functional form of the BCFT one‑point function for twist operators
\begin{equation}\label{eq:onept_mirrors}
\langle \sigma_n(u_i,v_i) \rangle \propto 
\left(\frac{\sqrt{p'(u_i)}}{v_i-p(u_i)}\right)^{\Delta_n},
\end{equation}
which can be deduced from \eqref{eq:onept_eucl} via the map~\eqref{eq:map_mirrors_bdy}.
The disconnected geodesic anchored on the brane reproduces the boundary‑channel contribution to the entropy.

\vspace{4pt}
\noindent\emph{Spacelike boundaries and complex geodesics.}\;
If the initial static boundary, which is related to the final state projection \cite{Akal:2021dqt}, is instead chosen to be spacelike, i.e.\ $\tilde{t}=0$ or $\tilde{u}=-\tilde{v}$, the conformal map returns $v = -p(u)$. This corresponds to the setup where the boundary of the BCFT is spacelike with non-trivial $t$-dependence.

On the gravity side, the disconnected geodesic length
\begin{equation}\label{eq:d_mirrors}
 D_{\mathrm{disc}}(u_i,v_i) = \sum_{i=1}^2 \left(\log\frac{|v_i + p(u_i)|}{\epsilon\sqrt{p'(u_i)}}\right) + i\pi\,
\end{equation}
is now complex.
When extremizing the length functional, as shown in the Supplemental Material, the geodesic is anchored to the mirror at
\begin{equation}
    z_I(u_i,v_i)=i\,\frac{p(u_i)+v_i}{2\sqrt{p'(u_I)}}\in \mathcal{B}_2(z\in \mathbb{C})\,,
\end{equation}
which is generically complex for any endpoint $A_i$.

On the field theory side, twist-operator one‑point functions are obtained from \eqref{eq:onept_mirrors}, now with $v_i=-p(u_i)$,
\begin{equation}
\langle \sigma_n(u_i,v_i) \rangle \propto 
\left(-\frac{\sqrt{p'(u_i)}}{\left|v_i + p(u_i)\right|}\right)^{\Delta_n},\quad i=1,2\,.
\end{equation}
It is important to observe that we can apply the same prescription for time ordering as in the case of static branes, since the causal structure is preserved by the conformal transformation \eqref{eq:map_mirrors_bdy}. After taking the logarithm as in \eqref{eq:ent_twist_correlator}, the entanglement pseudo entropy becomes
\begin{equation}
    S_A^{\mathrm{disc}}=\frac{c}{6}\sum_{i=1}^2 \left(\log\frac{|v_i + p(u_i)|}{\epsilon\sqrt{p'(u_i)}}\right) + i\frac{\pi c}{6}\ ,
\end{equation}
which matches the result from the bulk length \eqref{eq:d_mirrors}.

The agreement between the gravitational result and the field‑theoretic one confirms that the complexification of the EOW brane persists in time‑dependent setups, and that the resulting complex geodesic lengths faithfully capture the pseudo-entropy structure of the BCFT.
\\

\noindent
\textbf{\emph{Outlook.--}} The main result of this Letter is that systems described by holographic CFTs with boundaries of generic signature cannot be studied entirely within the holographic duality to a real spacetime. Holographically relevant bulk objects, like RT surfaces and EOW branes, extend to complex geometries by analytically continuing their coordinates. The resulting (pseudo) entanglement entropy is in general complex-valued and matches the boundary computation in terms of twist-field two-point functions, both in static and time-dependent scenarios.

The complex nature of the bulk objects emerges directly from extremizing the hypersurface area functional. Given the line element, our proposal can in principle be applied to any bulk geometry, also in higher dimensions. While the absence of an explicit field theoretical counterpart does not make these cases suitable for establishing the proposal, it is still relevant to exploit its bulk side to explore the properties of higher-dimensional islands, which can highly differ from lower-dimensional examples \cite{Heller:2024whi,Heller:2025kvp}.

From a more general perspective, our results motivate a precise understanding of the conditions under which the holographic duality to a complex spacetime can be deemed consistent. The most relevant criterion is the KSW condition for complex metrics \cite{Kontsevich:2021dmb,Witten:2021nzp}, which has been recently applied to holographic entanglement in complex spacetimes~\cite{Guo:2026vkl}. We expect that a similar analysis applied to the AdS/BCFT correspondence can provide a non-trivial test for such constructions. Given the relevance of complex metrics for timelike entanglement \cite{Heller:2024whi,Heller:2025kvp,Bernamonti:2026pxo}, dS holography \cite{Fujiki:2025rtx,Nanda:2025tid,Narayan:2026wzp} and more generally for dS \cite{Festuccia:2005pi,Fidkowski:2003nf,Chapman:2022mqd,Aalsma:2022eru} and black hole physics near spacetime singularities \cite{Ceplak:2024bja,Araya:2026shz}, this research direction can lead to diverse applications.

To conclude, tensor network simulations of a system with spacelike boundaries may be able to reproduce the entanglement (pseudo) entropy value and its phase transition as observed in holography~\cite{Carignano:2024jxb,Bou-Comas:2026gaa}. A promising starting point in this direction would be to extend the analysis of~\cite{Iino:2019vxd,Iino:2019rxt,Iino:2020ipa} to systems with spacelike boundaries.
\\

We would like to thank Wu-zhong Guo and Tadashi Takayanagi for useful discussions and comments on the draft.
FO thanks Alexey Milekhin for providing useful indications on free fermion simulations for temporal entanglement, and Bruno de Souza Le\~ao Torres for discussions on related topics. FO is supported by the Research Foundation Flanders (FWO) doctoral fellowship 1182825N. FO is grateful for the hospitality to the Yukawa Institute for Theoretical Physics in Kyoto, where some of the ideas inspiring this work emerged, and acknowledges financial support from the FWO grant for long stay abroad~V439125N.

\bibliographystyle{bibstyl}
\bibliography{biblio}

\clearpage

\appendix

\clearpage

\onecolumngrid

\setcounter{page}{1}
\setcounter{equation}{0}
\setcounter{figure}{0}
\renewcommand{\theequation}{S\arabic{equation}}
\renewcommand{\thefigure}{S\arabic{figure}}

\section{\large \mbox{Supplemental Material}}

\subsection{Complex branes from length functional extremization}

\noindent
In this section, we discuss in more detail the derivation of the complex nature of the brane in two instances of the AdS$_3$/BCFT$_2$ correspondence with a spacelike boundary, as presented in the main text. The two crucial conceptual ingredients are requiring that (1) the dual objects are extremal surfaces, and (2) time ordering between boundary and brane-anchored points is preserved when going from timelike to spacelike boundaries.

\subsubsection*{1.\ Constant-time brane in Poincar\'e AdS$_3$}

\begin{figure}
    \centering
    \includegraphics[width=.9\linewidth]{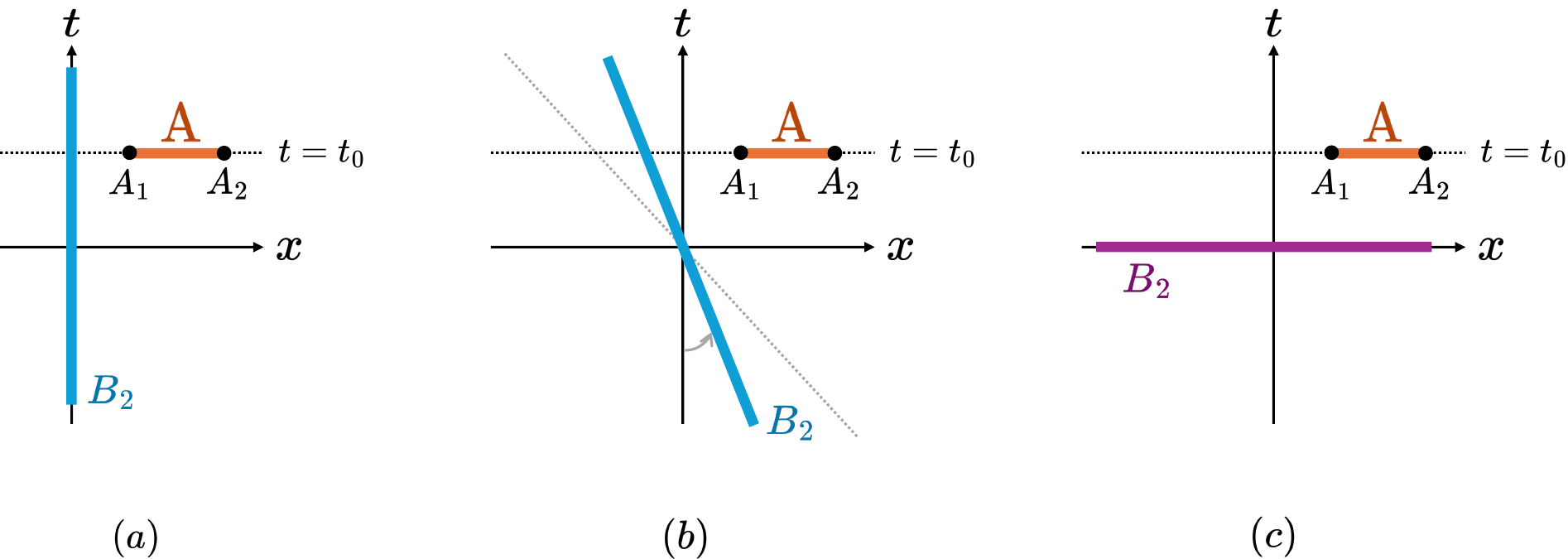}\vspace{.3cm}
    \caption{Pictorial representation of the analytic continuation from a timelike to a spacelike boundary $B_2$ in a BCFT$_2$. (\textit{a}) In the purely timelike case, there is no need to specify any time ordering between the endpoints and the boundary, since every significant correlation happens on the constant time-slice $t=t_0$. (\textit{b}) When the boundary is boosted and crosses the light-cone, an ordering prescription has to be specified, (\textit{c}) which then fixes the complex value of the entropy for spacelike boundaries.
    }
    \label{fig:rotation}
\end{figure}

\noindent
The metric is $ds^2 = z^{-2}(-dt^2+dx^2+dz^2)$ and $\mathcal{B}_2$ is the plane $t=0$. We consider a spacelike interval $A$ lying on the $t=t_0$ constant time-slice with endpoints $A_1=(t_1,x_1,z_1)=(t_0,x_1,\epsilon)$ and $A_2=(t_2,x_2,z_2) =(t_0,x_1+L,\epsilon)$, where $t_0>0$, $L>0$. In the disconnected phase, which is the phase of interest for studying the bulk brane, two geodesics connect the endpoints $A_1$, $A_2$ to points $I_1=(0,x_{I1},z_{I1})$ and $I_2=(0,x_{I2},z_{I2})$ on the brane. To establish the values of the coordinates of $I_1$, $I_2$ we extremize the length functional
\begin{equation}
    \mathrm{Len}=\int_\gamma d\lambda\ \sqrt{\frac{-\dot t^2(\lambda)+\dot x^2(\lambda)+\dot z^2(\lambda)}{z^2(\lambda)}}
\end{equation}
along a trajectory $\gamma:(t(\lambda),x(\lambda),z(\lambda))$ that satisfies the appropriate boundary conditions. In practical terms, the length of geodesics in AdS$_3$ is known in closed form given two generic endpoints $(t,x,z)$ and $(t',x',z')$:
\begin{equation}
    D(t,t',x,x',z,z')=\operatorname{arccosh}\left(-\frac{-(t-t')^2+(x-x')^2+z^2+z'^2}{2zz'}\right).
\end{equation}
For our purposes, the total length is then given by
\begin{align}
    D_\mathrm{disc} &=D(t_0,0,x_1,x_{I1},\epsilon,z_{I1})+D(t_0,0,x_1+L,x_{I2},\epsilon,z_{I2})\\[.5em]
    &= \log \left[\frac{-t_0^2+(x_1-x_{I1})^2+z_{I1}^2}{z_{I1} \epsilon }\right]+\log \left[\frac{-t_0^2+(x_1+L-x_{I2})^2+z_{I2}^2}{z_{I2} \epsilon }\right]+\mathcal{O}(\epsilon)\,,
\end{align}
where in the second line we expanded in the small regulator $\epsilon\to 0$. To extremize the length function we then require that its derivatives with respect to the four free parameters $x_{I1}$, $x_{I2}$, $z_{I1}$, $z_{I2}$ vanish. This always leads to $x_{I1}=x_1$ and $x_{I2}=x_2=x_1+L$, which could have been guessed by symmetry, and to four different possible imaginary solutions for the holographic coordinate,
\begin{equation}\label{eq:zPvals}
    \begin{cases}
        z_{I1}=\pm it_0\\
        z_{I2}=\pm it_0
    \end{cases}
    \ ,\qquad
    \begin{cases}
        z_{I1}=\pm it_0\\
        z_{I2}=\mp it_0
    \end{cases}
    .
\end{equation}
Adopting the logic employed in \cite{Bernamonti:2026pxo} for timelike entanglement entropy, it is easy to see that these correspond to the four possible time orderings of the endpoints $A_1$, $A_2$ with respect to the boundary of the BCFT (the former pair of solutions entailing the same time-ordering for both endpoints, while the latter opposite ones). The dual surfaces, indeed, are complex geodesics similar to the ones arising in the context of timelike entanglement entropy~\cite{Heller:2024whi,Heller:2025kvp}, but with only one boundary endpoint. They can be obtained from boosting the real, spacelike geodesics anchored on an interval of the same length (in absolute value) along the space direction, until their endpoints are separated in the time direction. This is equivalent to boosting the brane from timelike to spacelike signature while keeping the boundary points fixed, see Fig.~\ref{fig:rotation} for a visualization. Crossing the light-cone entails the choice of a regularization procedure and hence some multiplicity in the resulting entropy. We select the entanglement entropy configuration by requiring that the time ordering between the endpoints $A_1$, $A_2$ and the boundary is preserved along the boost.

To this end, let us focus on the first endpoint $A_1$ (the other being equivalent) and consider the length of the spacelike geodesic connecting $A_1=(t_1,x_1,\epsilon)$ to a point $I'_1=(t_{I'1},x_{I'1},z_{I'1})$ on the brane. Using the same formulae as before,
\begin{equation}
    D_1=\mathrm{dist}(t_1,t_{I'1},x_1,x_{I'1},\epsilon,z_{I'1})=\log\left[\frac{-(t_1-t_{I'1})^2+(x_1-x_{I'1})^2+z_{I'1}^2}{z_{I'1}\epsilon}\right].
\end{equation}
Following the ideas of \cite{Bernamonti:2026pxo}, we demand the endpoints to be time-ordered during the boost as we cross the light-cone, which is equivalent to employing the $i\varepsilon$-prescription $(t_1-t_{I'1})\to (t_1-t_{I'1})-i\varepsilon$. Then,
\begin{equation}
\begin{aligned}
    D_1 &=\lim_{\varepsilon\to 0}\log\left[\frac{-(t_1-t_{I'1}-i\varepsilon)^2+(x_1-x_{I'1})^2+z_{I'1}^2}{z_{I'1}\epsilon}\right]=\log\left[\frac{-(t_1-t_{I'1})^2+(x_1-x_{I'1})^2+z_{I'1}^2}{z_{I'1}\epsilon}\, e^{i\frac{\pi}{2}}\right]\\[.5em]
    &=\log\left|\frac{-(t_1-t_{I'1})^2+(x_1-x_{I'1})^2+z_{I'1}^2}{z_{I'1}\epsilon}\right|+i\frac{\pi}{2}
\end{aligned}
\end{equation}
when the separation between $A_1$ and $I_1'$ is timelike. Note that in evaluating the phase we kept in mind the extremal condition $z_{I'1}\propto (t_1-t_{I'1})$. When specifying to the points $A_1$ and $I_1$ above, we get the unique value for the length
\begin{equation}
    D_1=\log\frac{2t_0}{\epsilon}+i\frac{\pi}{2}\ ,
\end{equation}
which corresponds to $z_{I1}=it_0$ in \eqref{eq:zPvals}. The same procedure applied to the geodesic anchored to the other endpoint of the interval $A_2$ leads to the same result, so that the total length of the disconnected configuration is
\begin{equation}
    D_{\mathrm{disc}}=2D_1=2\log\frac{2t_0}{\epsilon}+i\pi\,.
\end{equation}
As discussed in the main text of the Letter, this result arises in a completely equivalent manner from the BCFT side, where time ordering fixes the same value for the imaginary part.

\subsubsection*{2.\ de Sitter brane with finite cutoff}

\noindent
The bulk metric is again empty AdS$_3$, but written in an AdS$_3$/dS$_2$ slicing,
\begin{align}
  ds^2 = d\eta^2 + \sinh^2\!\eta\,(-dt^2 + \cosh^2\!t\,d\theta^2)\,,
\end{align}
with the asymptotic boundary at $\eta=\eta_\infty\sim \log\frac{2}{\epsilon}$ and the brane $\mathcal{B}_2$ located at a finite radius $\eta=\eta_b$. We consider an interval with endpoints $A_1=(t_1,\theta_1,\eta_1)=(t_A,-\theta_A,\eta_\infty)$ and $A_2=(t_2,\theta_2,\eta_2)=(t_A,\theta_A,\eta_\infty)$, and focus again on the disconnected configuration. As before, it is convenient to use the closed form expression for the length of a geodesic connecting $A_1$ to a point $I_1=(t_I,\theta_I,\eta_b)$ on $\mathcal{B}_2$,
\begin{equation}
\cosh D_{AI} = \cosh\eta_\infty\cosh\eta_b - \sinh\eta_\infty\sinh\eta_b \left[\cosh t_A\cosh t_I\cos(\theta_I+\theta_A)-\sinh t_A\sinh t_I\right]\,.
\end{equation}
The same can be done for the other endpoint $A_2$. Extremization of the length functional over $t_I,\theta_I$ yields the family of complex saddles
\begin{align}\label{eq:app_complex_dS_saddles}
  \theta_{I1} = - \theta_A + \frac{\pi}{2} + \pi k\,,\qquad
  t_{I1} = i\pi\Bigl(m+\frac12\Bigr)\,,
\end{align}
where $k,m\in\mathbb{Z}$. The length of these saddles is also complex,
\begin{align}
  \cosh D_{AI} = \cosh\eta_\infty\cosh\eta_b + i(-1)^m \sinh\eta_\infty\sinh\eta_b \sinh t_A\,,
\end{align}
and in the limit $\eta_\infty\to \infty$,
\begin{equation}
    D_\mathrm{disc}=2D_{AI}=2 \eta_\infty +2\log\left[\cosh\eta_b + i(-1)^m \sinh\eta_b \sinh t_A\right].
\end{equation}
According to the proposed RT-like formula, the pseudo entropy will be given by the saddle with minimal real part:
\begin{equation}
    S_A^{\mathrm{disc}} = \min_{m\in \mathbb{Z}}\, \frac{c}{3}\left\{\eta_\infty+\log\left[\cosh\eta_b+ i(-1)^m \sinh\eta_b \sinh t_A\right]\right\},
\end{equation}
where the minimization is understood to act only on the real part.

There is an additional family of saddles,
\begin{equation}\label{eq:app_ds_realsaddles}
    \theta_{I1}=-\theta_A+\pi k\,,\qquad t_{I1}=\pm t_A+i\pi m\,,
\end{equation}
which instead give rise to a real length
\begin{equation}
    D_{AI}=\log\left[2\cosh(\eta_b\pm\eta_\infty)\right]\sim \eta_{\infty}\pm\eta_b\,.
\end{equation}
As discussed in \cite{Hao:2024nhd}, however, these saddles are unphysical, since (1) they are maximal with respect to variations along both the spatial and Euclidean time
directions and do
not correspond to the dominant saddle in the Euclidean path integral for calculating the
entanglement entropy, and (2) the corresponding real extremal surfaces intersect at the center of the spacetime and fail to meet the general topological conditions for RT surfaces. A quick way to see that (1) holds is to compute
\begin{equation}
\begin{aligned}
    \frac{\partial D_{AI}}{\partial \tau_I^2}\Biggr|_{\eqref{eq:app_ds_realsaddles}} =-\frac{\partial D_{AI}}{\partial t_I^2}\Biggr|_{\eqref{eq:app_ds_realsaddles}} &=-\sinh\eta_b \sinh \eta_\infty<0\\[.5em]
    \frac{\partial D_{AI}}{\partial \theta_I^2}\Biggr|_{\eqref{eq:app_ds_realsaddles}} &=-\cos^2 \tau_A\, \sinh{\eta_b}\,\sinh{\eta_\infty}<0
\end{aligned}
\end{equation}
and the determinant of the Hessian matrix $\sim\cos^2\tau_A\,\sinh^2\eta_b>0$ with respect to the variables $(\tau_I,\theta_I)$, where $\tau_I=it_I$ is the standard Euclidean time.

Both issues arising from (1) and (2) can be solved by considering instead the complex saddles \eqref{eq:app_complex_dS_saddles} and allowing the brane to extend to complex coordinates. As for (1), the eigenvalues of the Hessian matrix have opposite signs, since the matrix determinant is
\begin{equation}
    -\frac{1}{(\coth\eta_b\, \coth \eta_\infty\, \sec \tau_A+ \tan \tau_A)^2}<0\,,
\end{equation}
and hence we have a saddle point. Issue (2) does not arise in this case because the extremal surface is understood to live in the complexified spacetime and is smooth, with no self-intersections nor singular points.

\subsection*{More details on the island computation}

\noindent
In this appendix, we give a more detailed analysis of the island formula \eqref{eq:island_formula} in the case of a BCFT$_2$ coupled to a dS$_2$ spacetime with dynamical gravity discussed in the main text. Consider global dS$_2$ spacetime, which can be parametrized by $(t, \theta)$ coordinates with the metric
\begin{equation}\label{eq:global_dS}
ds^2 =L^2 \left(  - dt^2 + \cosh^2 t  \, d \theta^2 \right)   = \frac{L^2}{\cos^2 T} \left(  - dT^2 + d \theta^2 \right) \,,
\end{equation}
where in the last equality we applied the conformal coordinate transformation $\sinh t = \tan T$, $\cosh t = \frac{1}{\cos T}$ with a compact Lorentzian time $T \in (-\frac{\pi}{2}, \frac{\pi}{2})$. The spatial circle $S^1$ in the dS$_2$ metric has circumference equal to $2\pi L$, i.e.\ $ \theta\in[-\pi, \pi]$. As dS$_2$ spacetime is conformally flat, we have
\begin{equation}\label{eq:dSflat}
\quad ds^2 = \frac{L^2}{\Omega^2} \,dz\, d\bar{z} = \frac{4 L^2}{(1+ z\bar{z})^2}\, dz\, d\bar{z} \,,
\end{equation}
with a conformal factor $\Omega = \frac{1}{2} (1+z \bar{z})$. The Lorentzian global coordinates $(z, \bar{z})$ are then written as 
\begin{equation}\label{eq:zTtheta}
z = e^{-i (T -\theta)} \,, \qquad \bar{z} = e^{-i (T+\theta)}\,. 
\end{equation}

In the following, we will assume that dS$_2$ gravity is confined to a half circle, with angular coordinate
$\theta \in [-\frac{\pi}{2}, \frac{\pi}{2}]$, where we impose Neumann boundary conditions. For the other half of the circle, gravity is turned off by imposing Dirichlet boundary conditions.
For convenience, it is assumed that the dynamical matter on the full dS$_2$ background is still CFT$_2$. In other words, we glue a half dS$_2$ gravity with a CFT$_2$ bath system along a defect line at $\theta=\theta_0 = \pm\frac{\pi}{2}$.
We are interested in evaluating the entanglement entropy and its time evolution for a boundary interval $A$ located within the CFT$_2$ bath $A : \{ t = t_A \,,\, \theta \in [-\theta_A,\theta_A]  \}$.
It has been pointed out that the fine-grained entropy of a non-gravitating region entangled with a gravitational system is given by the so-called island formula \cite{Penington:2019npb,Almheiri:2019psf}
\begin{equation}
S_I(\hat{\rho}_{A}) = \min_I \operatorname*{ext}_{I} \left[ \frac{\mathrm{Area}(\partial I)}{4G_N} + S_{\mathrm{matter}}(A \cup I) \right],
\end{equation}
where the first area term denotes the gravitational entropy of the island $I$ with boundary $\partial I$, which in our case is a constant \cite{Hao:2024nhd}, while the second term is a bulk entropy and represents the quantum corrections given by the von Neumann entropy of matter fields.
The position of island $I$ is determined by its boundary $\partial I$, which is obtained by extremizing the generalized entropy. 

To derive the entanglement entropy of a single interval $A=[z_1,z_2]$ in the global dS coordinates we can use the flat-spacetime result
\begin{equation}
    S_\mathrm{flat}=\frac{c}{6}\log\left[\frac{(z_1-z_2)(\bar z_1-\bar z_2)}{\epsilon_1\epsilon_2}\right],
\end{equation}
where we have introduced two UV cutoffs $\epsilon_i$, $i=1,2$ associated to the endpoints, and count the appropriate Weyl factors~\eqref{eq:dSflat} for each endpoint of the interval, which amounts to shift the cutoffs $\epsilon_i\to \Omega_i\epsilon_i$, with $\Omega_i=\frac{1}{2}(1+z_i\bar z_i)$. Then, using \eqref{eq:zTtheta} and the conformal transformation to Lorentzian time $t$,
\begin{equation}\label{eq:bulkentropy}
     S_{\mathrm{matter}}(\theta_1,\theta_2) = \frac{c}{6} \log \left[ \frac{2(1+\sinh t_1 \sinh t_2- \cosh t_1 \cosh t_2 \cos(\theta_1 -\theta_2) ) }{\epsilon_1 \epsilon_2} \right] \,.
\end{equation}

In the non-island phase, the entanglement entropy of a single interval $A$ is given by
\begin{equation}\label{timecon}
 	S_I(\hat{\rho}_{A}) \big|_{\rm{non-island}}= S_{\rm matter}(A) = \frac{c}{3} \log \left( \frac{ 2\cosh t_A  \cdot  \sin\theta_A }{\epsilon}  \right),
\end{equation}
which presents a linear growth at late times,
\begin{equation}
S_I(\hat{\rho}_A) \big|_{\rm{non-island}} \sim \frac{c}{3} t_A \,. 
\end{equation}

Let us now consider the island phase, with the assumption that the boundary of the island region is located at $\partial I : \left\{  t=t_I, \theta = \pm \theta_I \right\}$.
Since the global Hartle-Hawking state is pure, the bulk entropy in the island phase can be calculated by considering the complementary region of $A \cup I$, which is the two-interval region $(A \cup I )^c$. In the semiclassical large-$c$ limit, we evaluate the two-interval entropy in the factorized channel, reducing the relevant twist correlator to two two-point functions with endpoints at $\theta=\pm\theta_A$ and $\theta=\pm\theta_I$. The corresponding pseudo entropy at the island phase can thus be obtained by applying the two-point function~\eqref{eq:bulkentropy},
\begin{align}
 S_I(\hat \rho_A) &= S_0 +  S_{\rm matter} (\theta_I, \theta_A)  + S_{\rm matter} (-\theta_A, -\theta_I) = S_0 + \frac{c}{3} \log \left[ \frac{2(1+\sinh t_I \sinh t_A- \cosh t_I \cosh t_A \cos(\theta_I -\theta_A) ) }{\epsilon \, \epsilon_I} \right] \nonumber\\[.5em]
 &= \tilde{S}_0 + \frac{c}{3} \log \left[ \frac{2(1+\sinh t_I \sinh t_A- \cosh t_I \cosh t_A \cos(\theta_I -\theta_A) ) }{\epsilon} \right],
\end{align}
where the cutoff scale at the island region $\epsilon_I$ has been absorbed into the constant term $S_0$ as $\tilde{S}_0=S_0+\frac{c}{3}\log\frac{1}{\epsilon_I}$. Extremization with respect to the anchorings $(t_I,\theta_I)$ is performed by solving 
\begin{equation}\label{solvbex}
    \begin{split}
  \partial_{\theta_I} S_I  =0 \quad &\longrightarrow \quad    \cosh t_I \cosh t_A \sin(\theta_I -\theta_A) =0 \,, \\
   \partial_{t_I} S_I  =0 \quad &\longrightarrow \quad   \cosh t_I \sinh t_A- \sinh t_I \cosh t_A \cos(\theta_I -\theta_A) =0 \,.
    \end{split}
\end{equation}
The above equations are solved by complex values for $t_I$, including the class of solutions
\begin{equation}
 \theta_I = \theta_A + \frac{\pi}{2} + k\pi\,,\qquad t_I = i\pi\left(m+\frac{1}{2}\right),\qquad k,m \in \mathbb{Z}\,,
\end{equation}
which lead to the final expression for the island formula in the disconnected phase
\begin{equation}
   S_I= \tilde{S}_0 + \frac{c}{3} \log \left[ \frac{2\left(1 + i (-1)^m \sinh t_A \right)}{\epsilon} \right].
\end{equation}
In the limit of large cutoff $\eta_b\to \infty$, this result matches \eqref{eq:S_disc_ds} up to a global additive constant, which can always be absorbed in the term $S_0$.

\subsection{Boundary entropy from complex branes}

\noindent
In this section, we show an example of how the boundary entropy term found from the BCFT side in \eqref{eq:Sdisc_bdy} can be interpreted geometrically, also in the case of branes extending in a complexified spacetime. To this end, let us consider the one‑parameter family of complex branes
$\mathcal{B}_2(\lambda):\; t = -i\lambda z$, $  \lambda\in\mathbb{R}$,
which includes the $t=0$ brane considered in the main text when $\lambda=0$. The brane is still anchored at the real boundary $t=z=0$.
Real values of $\lambda$ correspond to values of the tension $|\mathcal{T}|<1$, while the $|\mathcal{T}|>1$ case has been considered elsewhere and gives rise to real RT surfaces and branes \cite{Akal:2020wfl}. For an interval $A$ with endpoints $A_1=(t_0,x_1,\epsilon)$ and $A_2=(t_0,x_2=x_1+L,\epsilon)$, the disconnected channel is again represented by two geodesics joining $A_i$ to $I_i\in\mathcal{B}_2(\lambda)$, $i=1,2$.  
Using the brane condition $t_I=-i\lambda z_I$ and translational invariance in $x$, the distance formula gives
\begin{align}
\cosh D_i = 2+\frac{i\lambda t_0}{\epsilon} +\frac{1}{2\epsilon}\left[\frac{t_0^{2}}{z_I}-(1+\lambda^{2})z_I\right].
\end{align}
Extremization over $z_I$ yields the complex saddle
\begin{align}
z_I = \pm\frac{i t_0}{\sqrt{1+\lambda^{2}}}\ ,\qquad
t_I = \mp\frac{\lambda t_0}{\sqrt{1+\lambda^{2}}}\ ,
\end{align}
which again lies on the analytically continued brane $\mathcal{B}_2(\lambda;\,z\in\mathbb{C})$.
Inserting the branch that gives a positive imaginary part (by demanding time-ordering as in the $\lambda=0$ case) one finds
\begin{align}
D_i = \log \frac{2t_0}{\epsilon} + \log \bigl(\lambda+\sqrt{1+\lambda^{2}}\bigr) + i\frac{\pi}{2}.
\end{align}
Summing the two geodesics, the total disconnected length becomes $D_{\rm disc} =2D_i$.
The corresponding entanglement entropy $S_A^{\rm disc}=D_{\rm disc}/(4G_N)$ differs from the $\lambda=0$ result by the real boundary entropy
\begin{equation}
S_{\rm bdy}=\log g \sim\log\bigl(\lambda+\sqrt{1+\lambda^{2}}\bigr)
\end{equation}
that appears in the factorized boundary channel \eqref{eq:Sdisc_bdy}. The boundary entropy vanishes at $\lambda=0$, recovering the previous result. Note that the brane is complex in any case, independently of the value of $\lambda$.

\subsection{Entanglement entropy in Gaussian free fermionic systems with a spacelike boundary}

\noindent
In this section, we detail the computation of entanglement entropy for Gaussian free fermions, presented in the main text as a means of comparison to the same quantity in a BCFT with a static spacelike boundary. We will consider a (1+1)-dimensional theory of Dirac fermions, each of them with action
\begin{equation}
    S_\psi=\int dt\,dx\ \bar\psi \left(i\gamma^\mu\partial_\mu-m\right)\psi\,.
\end{equation}
In the following, we assume that the fermions are in a Gaussian state, i.e.\ such that the density matrix of the state is represented by a Gaussian function in fermionic phase space. A useful quantity to consider is the correlation matrix $C_{ij}=\braket{\bar \psi_i\psi_j}$, which contains all possible two-point functions between the different fermions. As we will briefly review, the correlation matrix allows one to compute the entanglement entropy of a subregion of the system. This well-known result \cite{Casini:2009sr} can be implemented numerically by discretizing the continuum theory onto a one-dimensional tight-binding lattice, where each lattice site hosts a fermion. We will consider the massless limit $m\to 0$: the dispersion relation is $\omega(k)=-\cos k$ and we can write each fermion in terms of creation and annihilation operators $a_k^\dagger$, $a_k$ as follows:
\begin{equation}
    \psi(t,x)=\int \frac{dk}{\sqrt{2\pi}}\ a_k\, e^{i(\omega(k)t+kx)}\,,\qquad
    \bar\psi(t,x)=\int \frac{dk}{\sqrt{2\pi}}\ a_k^\dagger\, e^{-i(\omega(k)t+kx)}\,.
\end{equation}
If we consider half-filled states, i.e.\ states occupied only in the region $|k|<\frac{\pi}{2}$, the two-point function $\braket{\bar\psi(t,x)\,\psi(0,0)}$ can be written as \cite{Milekhin:2025ycm}
\begin{equation}
    \braket{\bar\psi(t,x)\,\psi(0,0)} =\int_{-\frac{\pi}{2}}^{\frac{\pi}{2}} \frac{dk}{2\pi}\ e^{i(-t\cos k+kx)}
    = \frac{1}{4}\left[-it f\!\left(1,\frac{3-x}{2},\frac{3+x}{2},-\frac{t^2}{4}\right)+2f\!\left(1,\frac{2-x}{2},\frac{2+x}{2},-\frac{t^2}{4}\right)\right],
\end{equation}
where
\begin{equation}
    f(a,b_1,b_2,y)=\frac{_1F_2(a,b_1,b_2,y)}{\Gamma(b_1)\,\Gamma(b_2)}\,.
\end{equation}
The conjugate two-point function is just given by
\begin{equation}\label{eq:conjugation}
    \braket{\psi(t,x)\,\bar\psi(0,0)}=e^{i\pi x}\braket{\bar\psi(t,x)\,\psi(0,0)}\,.
\end{equation}
Note that we generically allow for fermions sitting on different time-slices. Once the correlation function $C_{ij}$ is built, its direct diagonalization in terms of a set of eigenvalues $\{\lambda_i\}_{i=1,\dots,N}$, leads to the entanglement entropy (or, more in general, pseudo-entropy) of the region $A$. If $N$ is the number of lattice sites included in the subregion, one has:
\begin{equation}\label{eq:ent_freeferm}
    S_A=-\sum_{i=1}^N \left[ \lambda_i\log \lambda_i+(1-\lambda_i)\log(1-\lambda_i)\right].
\end{equation}
For ordinary equal-time correlations this gives the usual Gaussian entanglement entropy. For the multi-time correlation matrix we are about to define, it instead implements the corresponding pseudo-entropy construction.

\begin{figure}[b]
    \centering
    \includegraphics[width=0.4\linewidth]{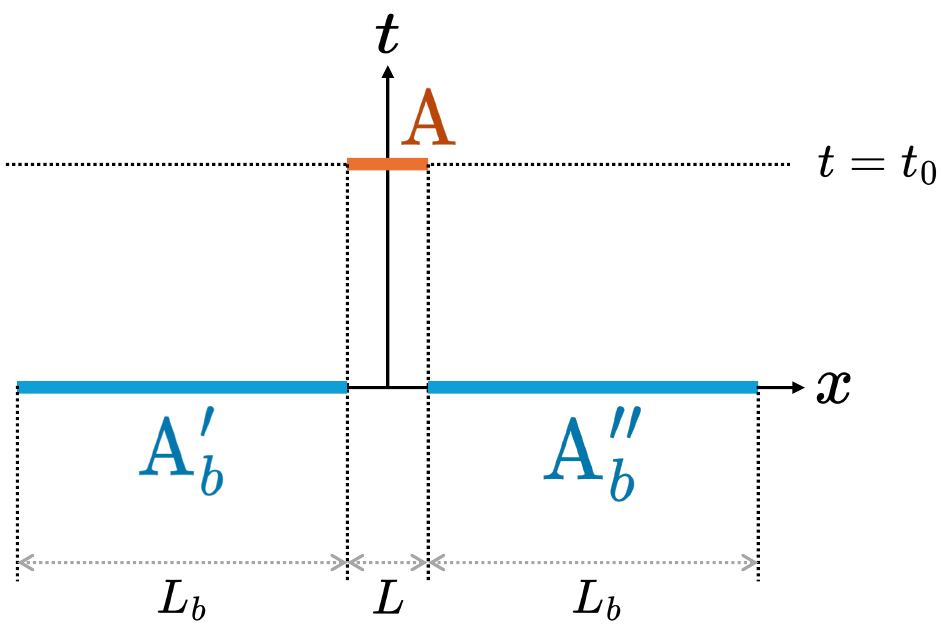}
    \caption{The three-interval configuration used to model a field theory with a spacelike boundary at $t=0$ in the context of free field computations. The entangling region $A$ lies on the time-slice $t=t_0>0$ and has length $L$ that is small compared to that of the regions $A'_b$ and $A''_b$ of length $L_b$ modelling the boundary, $L\ll L_b$. The continuum case is obtained asymptotically by sending $L_b/L$ to infinity.}
    \label{fig:free_fields}
\end{figure}

In practical terms, the way the correlation matrix is built depends on the geometry of the entangling regions. In the case of interest here, we consider a spacelike interval $A$ of size $L$ sitting on some constant-time slice $t=t_0>0$ and we model the presence of a spacelike boundary at $t=0$ by considering two semi-infinite regions $A_b'$, $A''_b$ at $t=0$, see Fig.~\ref{fig:free_fields}. This configuration is similar to the one proposed as a measure of temporal entanglement \cite{Milekhin:2025ycm}. Numerically, the size $L_b$ of the regions $A'_b$ and $A''_b$ has to be large enough compared to that of $A$, $L_b\gg L$, so that border effects coming from the finiteness of $L_b$ are negligible. The correlation matrix will then have a tripartite block structure,
\begin{equation}
    C_{ij}=\begin{pmatrix}
        C_{A'_bA'_b} & C_{A'_bA} & C_{A'_bA''_b}\\
        C_{AA'_b} & C_{AA} & C_{AA''_b}\\
        C_{A''_bA'_b} & C_{A''_bA} & C_{A''_bA''_b}
    \end{pmatrix},
\end{equation}
where each block contains correlations within the regions indicated in subscript. For example, the diagonal elements contain auto-correlations of a region within itself,
\begin{equation}
    C_{AA}=\begin{pmatrix}
        \braket{\bar\psi (t_0,x_1)\,\psi (t_0,x_1)} & \dots & \braket{\bar\psi (t_0,x_1)\,\psi (t_0,x_L)}\\
        \vdots & \ddots & \vdots\\
        \braket{\bar\psi (t_0,x_L)\,\psi (t_0,x_1)} & \dots & \braket{\bar\psi (t_0,x_L)\,\psi (t_0,x_L)}
    \end{pmatrix},
\end{equation}
while the off-diagonal elements contain correlations between different time-slices, such as
\begin{equation}
    C_{AA'_b}= \begin{pmatrix}
        \braket{\bar\psi (t_0,x_1)\,\psi (0,y_1)} & \dots & \braket{\bar\psi (t_0,x_1)\,\psi (0,y_{L_b})}\\
        \vdots & \ddots & \vdots\\
        \braket{\bar\psi (t_0,x_L)\,\psi (0,y_1)} & \dots & \braket{\bar\psi (t_0,x_L)\,\psi (0,y_{L_b})}
    \end{pmatrix},
\end{equation}
where for clarity we denoted the positions in the interval $A$ by $x_i$, $i=1,\dots ,L$, while in $A'_b$ by $y_j$, $j=1,\dots,L_b$. Note also that the off-diagonal elements are related by the conjugation \eqref{eq:conjugation}. The entanglement entropy is then evaluated by diagonalizing $C_{ij}$ as in \eqref{eq:ent_freeferm} and summing over \emph{all} its eigenvalues (and not only the ones relative to the region $A$, in analogy to the techniques developed in \cite{Milekhin:2025ycm}: in \eqref{eq:ent_freeferm}, then, $N=L+2L_b$). All the logarithmic functions are evaluated on their principal branch.

\begin{figure}[b]
    \centering
    \includegraphics[width=0.35\linewidth]{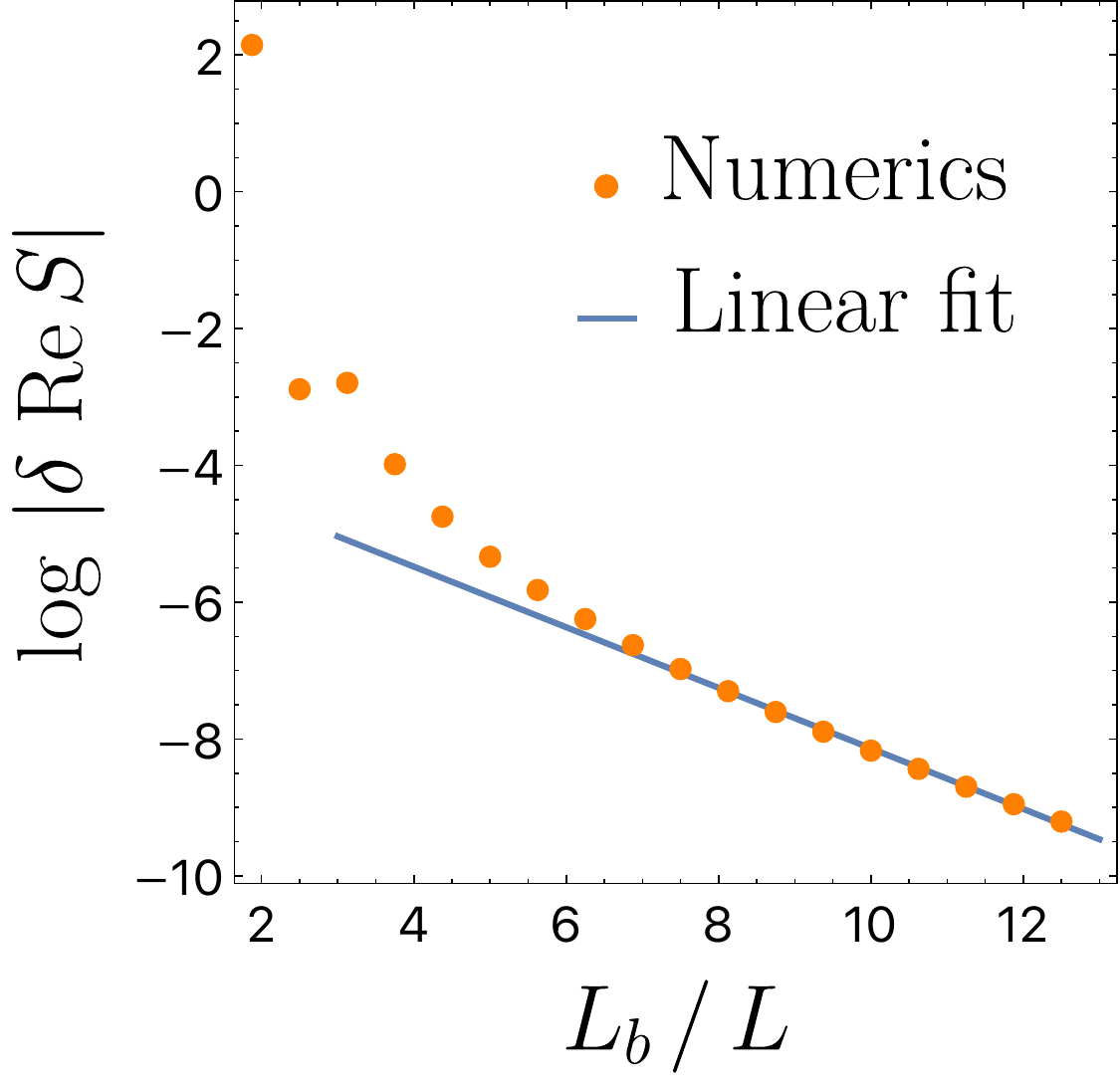}
    \hspace{.1\linewidth}
    \includegraphics[width=0.35\linewidth]{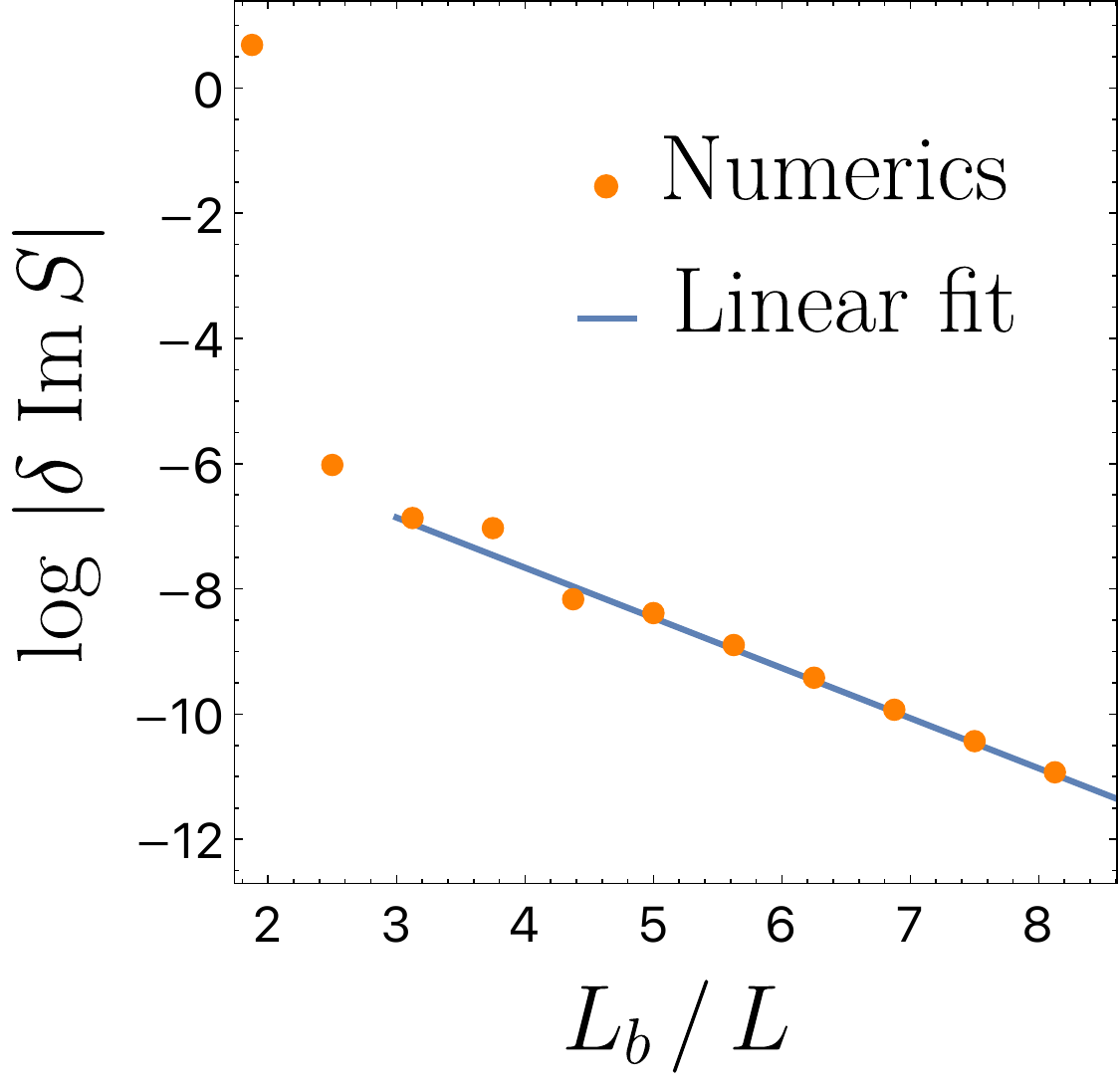}
    \caption{Convergence test for the real and imaginary part of the entanglement entropy as the boundary size $L_b$ is increased with a fixed interval size $L$. The vertical axes display the variations of the real and imaginary part of the entropy as the value of $L_b$ is varied \eqref{eq:varEnt}. In this figure the value $t_0=8$ has been used, the same as in Fig.~\ref{fig:con_t_ee}.}
    \label{fig:convergence}
\end{figure}

The plot in Fig.~\ref{fig:con_t_ee} of the main text has been obtained with $t_0=8$ and $L_b=300$, and $L$ ranges from $L=2$ to $L=48$. The match with holography holds up to an overall additive constant, which depends on the choice of normalization and can be fixed to be the same between the two cases. Moreover, one can check that the simulation converges to a finite value for the entropy as we take the limit $L_b\to \infty$ with a fixed value of $L$. This is shown in Fig.~\ref{fig:convergence}, where the variations in the real and imaginary part of the entropy,
\begin{equation}\label{eq:varEnt}
    \delta\operatorname{Re}S=\frac{\partial \operatorname{Re}S}{\partial L_b}\,\delta L_b\,,
    \qquad \delta\operatorname{Im}S=\frac{\partial \operatorname{Im}S}{\partial L_b}\,\delta L_b\,,
\end{equation}
are shown to decay exponentially as $L_b\to \infty$. These plots also show that the chosen values for $L$ and $L_b$ lie within the range of such exponential decay, making the relative variations due to finite boundary effects negligible.

\subsection{Complex saddles for moving mirrors}

\noindent
In this section, we discuss the disconnected geodesic saddles for moving mirror boundaries, emphasizing that the spacelike case generically forces the anchor point onto a complexified end‑of‑the‑world brane.
The bulk metric is
\begin{equation}
ds^2 = \frac{- du\,dv+dz^2}{z^2}\ ,
\end{equation}
and the mirror profile is $v = p(u)$ in the timelike case, or $v = -p(u)$ in the spacelike one, with $p'(u)>0$.

\subsubsection*{Static reference saddles}

\noindent
In the static frame $(\tilde u,\tilde v,\tilde z)$ the extremization of the length functional yields the following disconnected saddles for a boundary point $A=(\tilde u_A,\tilde v_A, \tilde z_A)$ connected to a brane point $I=(\tilde u_I, \tilde v_I, \tilde z_I)$:
\begin{align}
  \text{timelike brane:}\quad & \tilde u_I = \tilde v_I = \frac{\tilde u_A+\tilde v_A}{2}\ ,\quad 
    \tilde z_I = \frac{\tilde v_A-\tilde u_A}{2}\ , \label{st-t}\\[.5em]
  \text{spacelike brane:}\quad & \tilde u_I = \frac{\tilde u_A-\tilde v_A}{2}\ ,\quad 
    \tilde v_I = \frac{\tilde v_A-\tilde u_A}{2}\ ,\quad 
    \tilde z_I = i\,\frac{\tilde u_A+\tilde v_A}{2}\ . \label{st-s}
\end{align}
The spacelike saddle entails a complex brane, with $\tilde z_I$ purely imaginary.

\subsubsection*{Moving mirror from coordinate transformation}

\noindent
The chiral map $\tilde u = p(u)$, $\tilde v = v$ lifts to the bulk diffeomorphism \cite{Akal:2020twv}
\begin{equation}
  \tilde u = p(u)\,,\qquad 
  \tilde v = v + \frac{p''(u)}{2p'(u)}z^2\ ,\qquad 
  \tilde z = z\sqrt{p'(u)}\,. \label{bulk-trans}
\end{equation}
Applying the inverse transformation to the static saddles (\ref{st-t})--(\ref{st-s}) gives the anchor points in the presence of a moving mirror with trajectory specified by the function $v=p(u)$.
\\

\noindent\textit{Timelike mirror $v=p(u)$.}
For an endpoint $(u_A,v_A,z_A)$, the saddle satisfies
\begin{equation}
  p(u_I) = \frac{p(u_A)+v_A}{2}\,,\qquad
  v_I = p(u_I) - \frac{p''(u_I)}{2p'(u_I)}z_I^2\,,\qquad
  z_I = \frac{v_A-p(u_A)}{2\sqrt{p'(u_I)}}\ . \label{mov-t}
\end{equation}
The solution lies in the physical Poincaré patch only when $v_A > p(u_A)$, for which $z_I>0$.
\\

\noindent\textit{Spacelike mirror $v=-p(u)$.}
In this case we obtain
\begin{equation}
  p(u_I) = \frac{p(u_A)-v_A}{2}\ ,\qquad
  v_I = -p(u_I) - \frac{p''(u_I)}{2p'(u_I)}z_I^2\,,\qquad z_I = i\,\frac{v_A+p(u_A)}{2\sqrt{p'(u_I)}}\ . \label{mov-s}
\end{equation}
Since $z_I \propto i$, the saddle is generically complex, regardless of the endpoint. The corresponding geodesic length carries a universal imaginary part. Time ordering fixes its values in the same way as for static branes, since it is preserved by the conformal transformation $\tilde u=p(u)$, $\tilde v=v$ sending the static time $\tilde t=\tilde u+\tilde v$ to $t=u+v=p^{-1}(\tilde u)+\tilde v$. Then,
\begin{equation}
  D_1 = \log\left(\frac{|v_1+p(u_1)|}{\epsilon\sqrt{p'(u_1)}}\right) + \frac{i\pi}{2}. \label{length-s}
\end{equation}

Alternatively, one may also extremize the regulated length
\begin{equation}
\mathrm{dist}(u_I,v_I,z_I) = \log\left[\frac{z_I^2-(u_1-u_I)(v_1-v_I)}{\epsilon z_I}\right]
\end{equation}
subject to the bulk brane equation
\begin{equation}
v_I = \pm p(u_I) - \frac{p''(u_I)}{2p'(u_I)}z_I^2\,.
\end{equation}
The resulting algebraic conditions are identical to those leading to (\ref{mov-t}) and (\ref{mov-s}), confirming that the complex saddle is a genuine extremal point of the area functional.

\end{document}